\documentclass{aa}  

\usepackage[colorlinks=true, linkcolor = blue,
            urlcolor  = blue,
            citecolor = blue,
            anchorcolor = blue, bookmarks=false]{hyperref}
\usepackage{graphicx}
\usepackage{txfonts}
\usepackage{natbib}
\usepackage{natbib}
\usepackage{enumitem} 
\usepackage{color}
\usepackage{threeparttable}
\usepackage{tikz,amsmath}
\usetikzlibrary{shapes.geometric, arrows}
\usepackage{ulem}
\usepackage{wrapfig}
\usepackage{rotating}
\usepackage{lscape} 
\tikzstyle{process} = [rectangle, minimum width=1.5em, minimum height=3.5em, text centered, draw=blue, fill=gray!10]
\tikzstyle{process2} = [rectangle, minimum width=1.5em, minimum height=3.5em, text centered, draw=white, fill=white]
\tikzstyle{arrow} = [thick,->,>=stealth]
\usepackage{cancel}
\usepackage{ulem}
\bibpunct{(}{)}{;}{f}{}{,} 

\newcommand{\teff}{$T_{\mathrm{eff}}$}

\newcommand{\logg}{\mbox{log \textit{g}}}

\newcommand{\titans}{\textsc{Titans}}

\begin{document}

  \title{
   Thorium and Europium production in the early Galaxy 
    \thanks{Based on ESO archival spectra of the programmes 074.B-0639(A), 165.N-0276(A), 072.B-0585(A), 076.B-0055(A).}
}
\subtitle{Implications for nucleocosmochronology}

\author{L. Magrini\inst{1}
          \and
          R. E. Giribaldi\inst{1,2}
          \and
          M. Palla\inst{1}
          \and
          S. Randich\inst{1}
          \and
          L. Pasquini\inst{1}
          \and
          C. J. Hansen\inst{3}
          \and
          R. Smiljanic\inst{4}
          \and
          M. Racca\inst{5}
          \and
          M. Tsantaki\inst{1}
          \and
          J. Schiappacasse-Ulloa\inst{1,7}
          \and
          A. Skuladottir\inst{8}
          \and
          G. Kordopatis\inst{6}
          \and
          L. Berni\inst{1,8}
          \and
          T. Matsuno\inst{9}
          \and
          G. Casali\inst{10}
          \and
          V. Hill\inst{6}
          \and
          F. Matteucci\inst{11,12}
          \and
          G. Cescutti\inst{11,12}
          \and
          F. Rizzuti\inst{13}
          \and
          E. Spitoni\inst{12}
         }

\institute{
    INAF -- Osservatorio Astrofisico di Arcetri,
    Largo E. Fermi 5, 50125 Firenze, Italy\\
    \email{laura.magrini@inaf.it}
    \and
    Institute of Theoretical Physics and Astronomy,
    Faculty of Physics, Vilnius University,
    Sauletekio av. 3, 10257 Vilnius, Lithuania
    \and
    Institut f\"ur Theoretische Physik,
    Goethe-Universit\"at Frankfurt,
    Max-von-Laue-Str. 1, 60438 Frankfurt am Main, Germany
    \and
    Nicolaus Copernicus Astronomical Center,
    Polish Academy of Sciences,
    ul. Bartycka 18, 00-716 Warsaw, Poland
    \and
    Department of Astronomy, Stockholm University,
    Roslagstullsbacken 21, 114 21 Stockholm, Sweden
    \and
    Laboratoire Lagrange, Universit\'e C\^ote d'Azur,
    Observatoire de la C\^ote d'Azur, CNRS,
    Nice, France
    \and
    Leibniz Institute for Astrophysics Potsdam (AIP),
    An der Sternwarte 16, 14482 Potsdam, Germany
    \and
    Dipartimento di Fisica e Astronomia,
    Universit\`a degli Studi di Firenze,
    Via Sansone 1, 50019 Sesto Fiorentino, Italy
    \and
    Nicolaus Copernicus Astronomical Center, Polish Academy of Sciences,
    ul. Rabia\'nska 8, 87-100 Toru\'n, Poland
    \and
    Research School of Astronomy and Astrophysics,
    Australian National University,
    Canberra, ACT 2611, Australia
    \and
    Department of Physics, University of Trieste,
    Via Valerio 12, 34127 Trieste, Italy
    \and
    INAF -- Osservatorio Astronomico di Trieste,
    Via G. B. Tiepolo 11, 34143 Trieste, Italy
    \and
    Heidelberger Institut f\"ur Theoretische Studien,
    Schloss-Wolfsbrunnenweg 35, 69118 Heidelberg, Germany
    }           
    \date{Received October 16 2026;  accepted November 5 2026 }

 
  \abstract
{
The ages of the oldest stellar populations can be investigated through nucleocosmochronology. Among the various chronometers, the thorium-to-europium abundance ratio provides a promising indicator.
}
{
We investigate the reliability of the $\log({\rm Th}/{\rm Eu})$ ratio as a chronometer for old stellar populations by testing whether thorium and europium are co-produced in a constant ratio in metal-poor stars. We also assess the precision with which this abundance ratio can be measured and the resulting uncertainties in stellar ages, adopting the best available theoretical estimate of the initial production ratio to determine the absolute age scale.
}
{
We present new thorium and europium abundance measurements for a sample of eight \titans\, reference stars observed with UVES at VLT. We combine these measurements with homogeneous literature samples and analyse the evolution of $\log({\rm Th}/{\rm Eu})$ as a function of metallicity. Adopting a theoretical r-process production ratio, we derive ages and investigate the impact of observational uncertainties and sample statistics on the resulting age constraints.
}
{
In the low-metallicity regime, the observed $\log({\rm Th}/{\rm Eu})$ ratios show no significant dependence on metallicity once stars with anomalous actinide abundances are excluded, supporting the assumption that thorium and europium were typically co-produced  in the early Galaxy. The derived ages are broadly consistent with the expectation for a halo metal-poor population, although individual uncertainties remain large. We show that higher resolution is needed to  improve the precision of Th measurements, while the combination of several stars in coeval populations would provide an additional statistical improvement. 
}
{
Nucleocosmochronology  is a valuable independent method for dating the oldest stellar populations. Although individual stellar ages are currently limited by observational and theoretical uncertainties, the analysis of homogeneous samples of coeval stars, such as members of globular clusters, might provide significantly stronger constraints. Future high-resolution multi-object spectroscopic facilities will enable this approach and  provide independent constraints on the early formation history of the Milky Way and on the age of the Universe.
}


   \maketitle
%

\section{Introduction}

The study of heavy-element nucleosynthesis provides fundamental insights into the origin of the oldest stellar populations and the chemical evolution of the Galaxy \citep[see, e.g][]{Mishenina2022MNRAS.516.3786M}. Elements with $Z \geq 30$ are mainly produced through neutron-capture processes, among which the rapid neutron-capture process (r-process) is responsible for a significant fraction of the heavy-element inventory \citep{Burbidge1957RvMP...29..547B,Cameron1957PASP...69..201C}, particularly for the production of lanthanides ($Z = 57$--71) and actinides ($Z \geq 90$) \citep[e.g.][]{Goriely2001A&A...379.1113G}. Although neutron star mergers (NSMs) are currently the only observationally confirmed astrophysical sites of r-process nucleosynthesis \citep{Drout2017Sci...358.1570D, Cowperthwaite2017ApJ...848L..17C,   Watson2019Natur.574..497W,Domoto2025ApJ...978...99D}, other potential contributors, such as neutron star--black hole mergers, collapsars, and rare classes of core-collapse supernovae, are still under investigation \citep[see, e.g.,][]{Cowan2021RvMP...93a5002C, Siegel, Soker2026IJAA...16...94S}.
Observational evidence from metal-poor stars suggests that elements belonging to the second r-process peak, around A=130, and the third r-process peak, around A=195,  including the lanthanides, generally follow a nearly universal abundance pattern, consistent with production through a common r-process channel \citep[e.g.][]{Sneden1996ApJ...467..819S,Honda2006ApJ...643.1180H, Cowan2021RvMP...93a5002C}, but see also \citet{Puls2025} for the third r-process peak challenges. In contrast, the actinides show significant star-to-star variations, most notably in the so-called actinide-boost and actinide-deficient stars \citep[see, e.g.][]{hill2002A&A...387..560H,  Holmbeck2018ApJ...859L..24H}. These groups demonstrate that the production of the heaviest r-process nuclei depends sensitively on the physical conditions of the enrichment event, and it remains unclear whether such variations reflect multiple astrophysical sites or intrinsic diversity within a single class of r-process events \citep[see, e.g.,][]{Schatz2002ApJ...579..626S,Eichler2019ApJ...879...47E,    Farouqi2022A&A...663A..70F,  Wanajo2024PhRvL.133x1201W, Racca2025A&A...704A.282R, Jeong2026ApJ..1005...92J}. Despite this diversity, thorium (Th) often broadly follows the abundance pattern of the lighter r-process elements, particularly those associated with the second and third r-process peaks, such as europium (Eu). Among metal-poor r-process-enhanced stars, approximately 15\% exhibit an actinide boost \citep[see, e.g.,][]{Shah2026arXiv260412892S}. However, they represent only a small fraction of the overall metal-poor stellar population. Both estimates are affected by strong observational biases, particularly those associated with the detectability of Th lines \citep[e.g.][]{hill2002A&A...387..560H,Roederer2009ApJ...698.1963R}.

Radioactive nucleocosmochronology provides indeed a unique tool for estimating the ages of old stars, although its accuracy is currently limited by observational uncertainties and systematic effects associated with the assumed r-process Production Ratios \citep[PR,][]{Fowler1960AnPhy..10..280F,Cowan1987ApJ...323..543C,Cowan1991PhR...208..267C,Cowan1997ApJ...480..246C,Truran2001ASPC..245..226T,Sneden2001Natur.409..673S, Ludwig2010A&A...509A..84L}. 
The radioactive nature of Th makes it particularly valuable for stellar-age estimates. Radioactive nucleocosmochronology uses the decay of long-lived radioactive nuclei to constrain the ages of old stars, with the Th/Eu ratio providing one of the most widely used chronometers \citep{Fowler1960AnPhy..10..280F,Cowan1987ApJ...323..543C,Cowan1991PhR...208..267C,Truran2001ASPC..245..226T,Sneden2001Natur.409..673S}. Thorium is a long-lived radioactive actinide whose abundance decreases through radioactive decay, whereas Eu ($Z=63$), a lanthanide, is effectively stable on stellar-evolution timescales. Consequently, the present-day Th/Eu ratio retains information about the time elapsed since the production of these elements. In principle, the stellar age can therefore be inferred by comparing the observed ratio with the initial Th/Eu PR predicted by r-process nucleosynthesis models \citep{Frebel2007ApJ...660L.117F}. In this context, the $\log({\rm Th}/{\rm Eu})$ ratio can provide a reliable age indicator for old stellar populations in the Milky Way and its nearby companion galaxies \citep[e.g. in Sagittarius;][]{Hansen2018}, provided that stars with anomalous actinide-to-lanthanide ratios (i.e., actinide-boost/deficient stars) are excluded.

The main limitation of the Th/Eu chronometer is the uncertainty in this initial PR. Different r-process models predict a range of initial $\log(\mathrm{Th/Eu})$ values, with values in the literature spanning approximately $-0.34$ to $-0.24$ dex \citep{Cowan1999ApJ...521..194C,Kratz2007PrPNP..59..147K,Hill2017A&A...607A..91H,Placco2017ApJ...844...18P,Ji2018ApJ...856..138J}. This corresponds to an initial [Th/Eu] range of approximately 0.12--0.26 dex when adopting a solar reference value of $\log(\mathrm{Th/Eu})_\odot\simeq-0.50$ dex \citep{asplund2009ARA&A..47..481A}. Such differences in the assumed PR directly translate into systematic uncertainties in the derived stellar ages, in addition to observational uncertainties in the measured abundances \citep{Cayrel2001Natur.409..691C,Frebel2007ApJ...660L.117F}.

An additional consideration is whether Th and Eu are produced in a sufficiently consistent manner across different r-process events. Recent studies using stars with independent asteroseismic ages suggest possible variations in their relative production across Galactic epochs \citep{Azhari2025A&A...699A.276A}.

While in this paper we focus on Th, we mention that other abundance ratios (e.g., [Th/U] or [Th/Hf]) can be used as chronometers, with some advantages over [Th/Eu] in terms of the underlying production yields \citep[e.g.][]{Cowan1999ApJ...521..194C, Schatz2002ApJ...579..626S,Kratz2007ApJ...662...39K, frebel2018ARNPS..68..237F}. However, their use is severely limited by observational difficulties and theoretical uncertainties.

In this work, we expand the sample of metal-poor stars with measured Th and Eu abundances to investigate whether the relative production of Th and Eu is sufficiently stable across the metal-poor regime to justify the adoption of a common PR  and to evaluate the reliability of $\log({\rm Th}/{\rm Eu})$ as an age indicator.
The data sample and spectral analysis are described in Sections~\ref{sec:data} and \ref{sec:analysis}, respectively. In Sect.~\ref{sec:analysis}, we present also  our results, combined with literature samples. In Sect.~\ref{sec:PR}, we evaluate the Th/Eu PR as a function of [Fe/H] to investigate whether Th and Eu can be attributed to a common r-process origin and to quantify possible variations in their production. In Sect.~\ref{sec:ages}, we estimate the ages of our sample stars and of literature stars, exploring possible approaches to improve the applicability of nucleocosmochronology, based both on very-high-resolution spectroscopy and on the statistical advantages offered by coeval stellar populations, such as globular clusters. Finally, in Sect.~\ref{sec:summary}, we summarize our results and present our conclusions.

\section{Data sample}
\label{sec:data}
The \titans\ sample \citep{giribaldi2021A&A...650A.194G,giribaldi2023A&A...679A.110G} provides valuable benchmark stars in the metal-poor regime owing to the high precision of their stellar-parameter determinations, making them valuable reference objects for studies of stellar nucleocosmochronology. 
Their effective temperatures (\teff) have been demonstrated to be accurate to 50~K, with most of this uncertainty arising from the uncertainties of the reference fundamental values. Since these reference uncertainties dominate the accuracy assessment, the precision of the \teff\ of the \titans\ may be adopted as the total uncertainty, under an optimistic hypothesis.
This is one of the key contributions of the \titans\ sample. The high accuracy and precision of their \teff\ determinations optimise the precision of the chemical abundances, such that the remaining abundance uncertainties closely approach the fidelity of the line modelling.

Our sample is composed of the RGB \titans\, reference stars \citep{giribaldi2023A&A...679A.110G}, for which we used archival ESO UVES \citep{dekker2000} spectra obtained at a resolving power of $R = \lambda/\Delta\lambda \sim 40,000$. The spectra provide broad wavelength coverage, extending from $3340 \lesssim \lambda \lesssim 4580$~\AA\ in the blue and from $4800 \lesssim \lambda \lesssim 6800$~\AA\ in the red. 
A Carbon-Enhanced Metal-Poor (CEMP) star was excluded from the subsequent analysis, but its abundance measurements were retained to enable comparisons with previous literature studies.
We selected from the sample of \citet{giribaldi2023A&A...679A.110G} only stars showing a detectable Th feature at 4019.129~\AA, 
which is among the most commonly used and reliable lines for thorium abundance 
determinations \citep[see][for a list of Th features]{Gamrath2020MNRAS.496.4507G}. 
This selection was based on the quality of the available spectra, requiring a minimum signal-to-noise ratio (S/N) of 70 at about 402~nm and a minimum spectral resolution of $R = 40{,}970$. The sample of stars in common with the literature is listed in Table~\ref{tab:Th_comparison}, whereas the full sample is presented in Table~\ref{tab:parameters}, which also provides the resolution of the blue spectra, the number of co-added exposures, and the resulting S/N.

\section{Spectral analysis}
\label{sec:analysis}

\subsection{Atmospheric parameters}
The atmospheric parameters of our sample stars are listed in Table~\ref{tab:parameters}. These were derived  robustly,  starting from the parameters in \cite{giribaldi2023A&A...679A.110G},  as described in \cite{giribaldi2026A&A...712A.141G}.
Effective temperatures were determined by averaging the values obtained from H$\alpha$ profile fitting based on 3D NLTE models \citep{amarsi2018} and from the Infrared Flux Method \citep[IRFM,][]{Casagrande2010,casagrande2021}. The only exception is BD+11~2998, for which the H$\alpha$-based \teff\ was adopted because the two determinations differ by more than 1$\sigma$.

Metallicity ([Fe/H]), microturbulence ($v_{mic}$), and \logg\ are based on the ionization and excitation equilibrium of Fe lines under non-LTE, for which we used the departure coefficients based on the model atoms developed by \citet{bergemann2012MNRAS.427...27B} and \citet{semenova2020A&A...643A.164S}. We proceeded as follows, line-by-line analyses were done with 
synthetic spectra generated from fixed \teff\ using MARCS model atmospheres \citep{gustafson2008} running the Turbospectrum \citep{gerber2023} radiative transfer code.
The projected rotational velocity ($v\sin i$) was fixed at 0~km\,s$^{-1}$, while the macroturbulent velocity ($v_{\mathrm{mac}}$) was determined by minimising the difference between the observed and synthetic spectra. Consequently, the fitted macroturbulent broadening effectively accounts for both macroturbulence and any unresolved rotational broadening.

The \logg\ was derived by  ionisation balance, and it was validated by the theoretical relation, applied as described in Appendix~\ref{sec:logg}.
The latter method yields  slightly lower values than the ionization equilibrium ones (differences are indicated in brackets in Table~\ref{tab:parameters}). We estimated the uncertainty of \logg, 0.14~dex, by adding in quadrature the offset and the dispersion of the differences.
Our grids of non-LTE coefficients are incomplete around the parameters of the stars BPS~BS~17569-049 and HD~45282. Therefore, for them, \logg\ was derived solely from Eq.~\ref{eq:fundamental}. For the metallicity of these two stars to be compatible with the rest, their [Fe/H] was derived from \ion{Fe}{ii} lines, keeping \logg\ fixed, as the non-LTE corrections are negligible for \ion{Fe}{ii} \citep[e.g.][]{amarsi2016MNRAS.463.1518A,giribaldi2023A&A...679A.110G}.

For every star, microturbulence was optimised by averaging the results from these three methods: (i) balance of the reduced EW under 1D NLTE, when available, (ii) global line fitting\footnote{under 1D~LTE, via minimization of $\chi^2$ while keeping all atmospheric parameters fixed, including $v_{sin\,\rm{i}} = 0$~km/s, except $v_{\mathrm{mac}}$.}, and (iii) 3D~LTE calibrations \citep{dutra-ferreira2016A&A...585A..75D}. The uncertainty was estimated as the standard deviation of the three resulting values.

\subsection{Thorium abundance}

The Th~{\sc ii} feature at 4019.129~\AA\ is blended with atomic lines of Fe, Ce, Ni, and Co, and with molecular lines of $^{12}$CH and $^{13}$CH \citep[e.g.][]{Hansen2018,gull2018ApJ...862..174G, giribaldi2026A&A...711A..47G}. 
In addition, an unidentified feature  is observable at 4019.3~\AA\footnote{We refer to Alencastro Puls in prep. for updated atomic physics of the blends.}.
Therefore, prior to determining the Th abundance, we derive the abundances of the mentioned elements.

The carbon abundance was derived by fitting the molecular band at 4265--4307~\AA. Its uncertainty was determined by adding in quadrature the contributions from spectral noise, \teff, \logg, and $v_{mic}$. Each source contributes by approximately 0.02~dex, yielding a total abundance uncertainty of 0.04~dex.
We derived the $^{12}$C/$^{13}$C ratio by fitting the feature at 4213--4250~\AA.
Nickel, cerium, and cobalt abundances were determined by fitting the lines listed in \cite{giribaldi2026A&A...711A..47G}. The internal uncertainties were estimated from the line-to-line dispersion, whereas the total uncertainties also include the systematic effects of the errors in \teff, \logg, and $v_{mic}$, since the uncertainties in [Fe/H] are negligible.
For Ni and Co, the error budget is dominated by the uncertainty in \teff, which contributes about 0.04~dex, whereas for Ce both the uncertainties in \teff\ and \logg\ together contribute  about 0.05~dex.
To simulate the flux contribution of the unknown feature at 4019.3~\AA, a Gaussian function is fitted in a step prior to the line synthesis. 
With the exception of the CEMP star, the stars in our sample generally exhibit subsolar [Ni/Fe] ratios (median = $-0.16$~dex) and slightly supersolar [Ce/Fe] ratios (median = $+0.12$~dex), while Co remains undetected in most cases. The median $^{12}$C/$^{13}$C ratio is 95. Among the species contributing to the blended spectral feature, Fe and $^{13}$CH provide the dominant contributions. Nevertheless, Ni, Ce, and the unidentified feature may also introduce significant uncertainties in the derived Th abundance if their contributions are not accurately modelled.
The Th abundance was determined by fitting synthetic spectra to the Th~{\sc ii} feature at 4019.13~\AA, allowing the macroturbulent velocity ($v_{\mathrm{mac}}$) to vary freely. The best-fitting $v_{\mathrm{mac}}$ values agree with those derived from Fe lines to within $\pm1$~km\,s$^{-1}$, with a typical value of approximately 4.5~km\,s$^{-1}$.
Our Th abundances are listed in Table~\ref{tab:Eu_Th}, whereas details on the feature fits and the abundance error analysis are provided in Appendix~\ref{sec:th_profiles_err}.
We do not apply NLTE corrections to the Th abundances. Although \citet{mashonkina2012A&A...540A..98M} reported positive 1D NLTE corrections for Th~{\sc ii} lines in metal-poor giants (for our stars ranging from +0.03 to +0.09), their magnitude depends strongly on the adopted atomic model and treatment of collisions with hydrogen. Given these uncertainties, we retain a homogeneous 1D LTE analysis of Th throughout the sample to avoid introducing model-dependent corrections. Furthermore, \citet{caffau2008A&A...483..591C} highlighted the importance of 3D effects and line blending in the solar Th abundance determination. A fully consistent 3D NLTE treatment is ultimately required to assess the systematic uncertainties in Th abundances.

\subsection{Europium abundance}

For europium, the situation differs from that of thorium, as the Eu~{\sc ii} lines are generally stronger and NLTE effects can be evaluated more reliably. Four Eu~{\sc ii} transitions are covered by our spectra, at 4129.72, 4205.03, 6437.64, and 6645~\AA, although not all are detectable across the full metallicity range. The two bluest lines remain accessible at the lowest metallicities, whereas the redder transitions become increasingly difficult to measure.

NLTE calculations are available for the two bluest transitions \citep{storm2024}, while the 6437.64 and 6645~\AA\ lines are analysed under 1D LTE assumptions. We therefore derived Eu abundances from the blue lines using both LTE and NLTE line formation and compared them with the LTE abundances obtained from the redder lines.

For the 4129.72 and 4205.03~\AA\ transitions, we adopted the hyperfine structure (HFS) and atomic data from \citet{da_silva2025A&A...696A.122D}. Individual Eu abundances are reported in Table~\ref{tab:abundances} and compared in Fig.~\ref{fig:theu_comp}. Where all four transitions are detectable, we find good agreement between the LTE abundances from the redder lines and the NLTE abundances from the bluer lines. The resulting NLTE corrections are generally modest ($\lesssim0.08$ dex), indicating a limited impact on the derived Eu abundances.

Our final Eu abundances combine all available Eu~{\sc ii} lines, adopting NLTE departure coefficients from \citet{storm2024} for the two bluest transitions and retaining LTE for the redder lines. For HD 45282 and BPS CS 22892-052, whose atmospheric parameters fall outside the adopted NLTE grid, Eu abundances were retained under LTE assumptions.

The NLTE corrections adopted here are generally smaller than those predicted by the earlier calculations of \citet{mashonkina2012A&A...540A..98M,Mashonkina2014A&A...569A..43M}. As a result, our combination of LTE Th and NLTE Eu abundances yields $\log(\mathrm{Th/Eu})$ ratios broadly comparable to those estimated by applying the NLTE corrections of these authors to both elements, owing to the partial compensation of their effects. Excluding the lowest-gravity star, BPS BS 17569-049, the two approaches differ by only 0.019 dex on average (mean absolute difference of 0.033 dex). 
The error budget is dominated by the uncertainties in \teff\ and \logg. Their combined contribution added in quadrature is reported in Table~\ref{tab:parameters}.

\subsection{Comparison with literature results}

As a consistency check of our abundance analysis, we compared our A(Th), A(Eu), and $\log(\mathrm{Th/Eu})$ with previous literature determinations for the stars in common. As summarized in Table~\ref{tab:Th_comparison}, literature measurements are available for three stars. A comparison of the stellar parameters reveals a systematic offset in effective temperature, with our $T_{\rm eff}$ values being higher by approximately 200 K. Our effective temperatures are derived from 3D non-LTE modelling of the H$\alpha$ line profile. This temperature scale is known to differ systematically from temperatures obtained by enforcing excitation equilibrium of Fe lines, which tend to be underestimated in metal-poor giants \citep{mucciarelli2011A&A...528A..44M}. In contrast, the H$\alpha$ temperatures are in excellent agreement with independent temperature determinations from the IRFM and, where available, interferometry \citep{giribaldi2023A&A...679A.110G}. This provides strong support for the adopted \teff\ scale, which is particularly important given the strong sensitivity of the Th abundance to \teff, especially for stars approaching the tip of the RGB \citep{giribaldi2019A&A...624A..10G}. However, the difference in temperature with respect to the literature values has an effect on the abundances, which is comparable for both Th and Eu abundances, which are typically higher by about 0.2 dex than the literature values. Since the two elements are affected in a similar way, these systematic differences largely cancel out in the abundance ratio. Consequently, our $\log(\mathrm{Th/Eu})$ ratios are in very good agreement with previous determinations, supporting the robustness of our relative abundance analysis.

\begin{table*}
\centering
\tiny
\caption{Literature  stellar parameters, europium and thorium abundances,
and $\log(\mathrm{Th/Eu})$ ratios 
for stars in common with the present work. }
\label{tab:Th_comparison}
\begin{tabular}{lcccccc}
\hline
Star &
$T_{\rm eff}^{\rm lit.}$ &
$\log g^{\rm lit.}$ &
$[\mathrm{Fe/H}]^{\rm lit.}$ &
$A(\mathrm{Eu})^{\rm lit.}$ &
$A(\mathrm{Th})^{\rm lit.}$ &
$\log(\mathrm{Th/Eu})^{\rm lit.}$  \\
 &
[K] &
[dex] &
[dex] &
[dex] &
[dex] &
[dex]  \\
\hline
BD+17~3248$^{a}$ &
$5200\pm150$ &
$1.80\pm0.30$ &
$-2.0\pm0.2$ &
$-0.67\pm0.05$ &
$-1.18\pm0.10$ &
$-0.51\pm0.11$  \\

HD~186478$^{b}$ &
$4600$ &
$1.45$ &
$-2.50$ &
$-1.56\pm0.06$ &
$-2.26\pm0.07$ &
$-0.70\pm0.09$  \\


BPS CS~22892$-$052$^{c}$ &
$4800$ &
$1.50$ &
$-3.12$ &
$-0.95\pm0.03$ &
$-1.57\pm0.10$ &
$-0.62\pm0.10$ \\
\hline
\end{tabular}

\tablefoot{
$^{a}$ \citet{Cowan2002ApJ...572..861C};
$^{b}$ \citet{Johnson2001ApJ...554..888J};
$^{c}$ \citet{Sneden2003ApJ...591..936S}.
}
\end{table*}

\begin{figure}
    \centering
\includegraphics[width=1\linewidth]{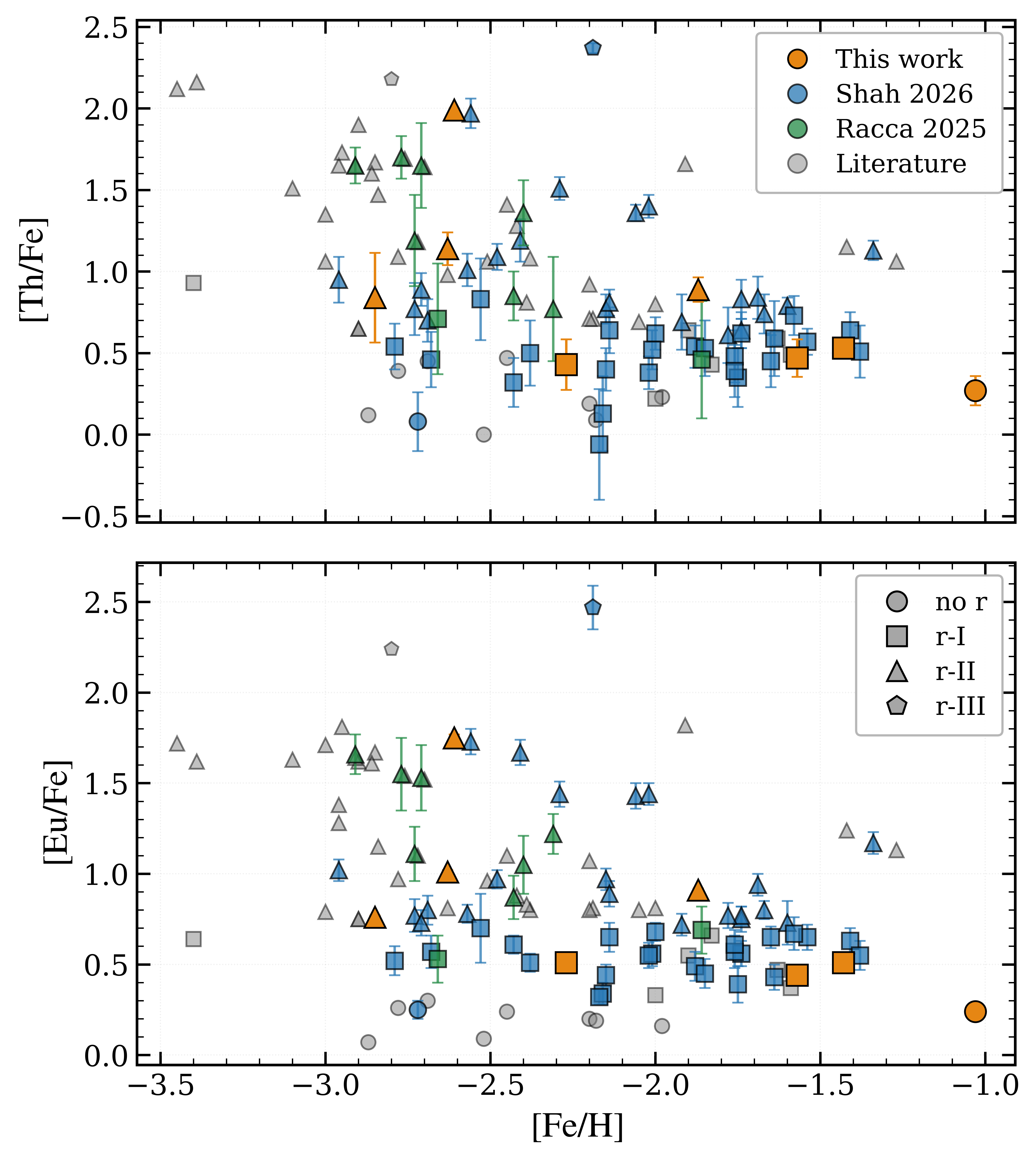}
    \caption{Upper panel: [Th/Eu] as a function of metallicity [Fe/H]; lower panel [Eu/Fe] versus [Fe/H]. The sample of the present work is shown with orange symbols, whereas the samples of \citet{Racca2025A&A...704A.282R} and \citet{Shah2026arXiv260412892S} are in green and blue, respectively. The literature sample collected in S26 is in transparent gray. The different shapes of the symbols indicate the r-process type star: non-RPE stars (circles)  are defined as having $[\mathrm{Eu}/\mathrm{Fe}] \leq 0.3$. r-I stars (squares) have $+0.3 < [\mathrm{Eu}/\mathrm{Fe}] \leq +0.7$, r-II stars (triangles) have $+0.7 < [\mathrm{Eu}/\mathrm{Fe}] \leq +2.0$, and r-III stars (pentagons) have $[\mathrm{Eu}/\mathrm{Fe}] > +2.0$.  }
    \label{fig:thfe_eufe}
\end{figure}

\subsection{Abundance trends}
Our results are presented in Table~\ref{tab:parameters} and in 
Fig.~\ref{fig:thfe_eufe}, where we also include the results from two recent 
studies in which Th and Eu abundances have been measured in metal-poor stars \citep[][hereafter R25 and S26, respectively]{Racca2025A&A...704A.282R, Shah2026arXiv260412892S}. 
We excluded the CEMP-r star 2MASS J20093393-3410273 from R25. 
Th and Eu abundances are also available for other large samples of stars \citep{Mishenina2013A&A...552A.128M, Mishenina2022MNRAS.516.3786M}. However, these stars lie in a substantially higher metallicity range and are therefore not included in our comparison sample.
Both our targets and the literature samples are classified into the different classes of r-process-enhanced (RPE) stars, following the scheme introduced by \citet{Holmbeck2020ApJS..249...30H}, as reported in the caption of Fig.~\ref{fig:thfe_eufe}.


We observe a predominance of non-r-process-enhanced (non-RPE; [Eu/Fe] $\leq +0.3$) and r-I ($+0.3 <$ [Eu/Fe] $\leq +0.7$) stars at higher metallicities, whereas r-II ($+0.7 <$ [Eu/Fe] $\leq +2.0$) and r-III ([Eu/Fe] $> +2.0$) stars become more frequent toward lower metallicities. 
This trend is likely related to the reduced number of enrichment events and the 
less efficient mixing of r-process products in the early Galactic halo. 
At a given metallicity, the average levels of [Th/Fe] and [Eu/Fe] increase toward lower metallicity, 
indicating that these neutron-capture elements were produced efficiently and on shorter time scales
relative to iron during the early stages of Galactic chemical evolution.  However, this trend should be interpreted with caution, as at [Fe/H] < -2 the detection of weak Th lines and low Eu abundances becomes increasingly challenging, introducing a selection bias toward stars with higher neutron-capture abundances, particularly since our sample includes only stars with detectable Th lines. 
The large scatter observed at low metallicity is consistent with the stochastic 
nature of r-process enrichment in the Galactic halo, as discussed by, e.g.,  
\citet{Cescutti2008A&A...481..691C, Hansen2012},  \citet{Cescutti2014A&A...565A..51C}, and \citet{Rizzuti2019MNRAS.489.5244R}.

\begin{table*}
\caption{Parameters and observation setups of our sample stars.}
\label{tab:parameters}
\centering
\tiny
\begin{threeparttable}
\begin{tabular}{l|ccccccc}
\hline\hline
Star & \teff\ & \logg\ & [Fe/H] & $v_{mic}$ & S/N & N & $\lambda/\Delta \lambda$\\
 & [K] & [dex] & [dex] & [km/s] & & & \\
\hline

BD+09 2860 & $5415\pm16$ & 2.30[$-0.11$] & $-1.57\pm0.07$ & $1.45\pm0.3$ & 231 & 2 & 40970 \\

BD+11 2998 & $5607\pm50$ & 2.73[$-0.10$] & $-1.03\pm0.09$ & $1.64\pm0.01$ & 286 & 1 & 40970 \\

BD+17 3248 & $5324\pm17$ & 2.15[$-0.07$] & $-1.87\pm0.07$ & $1.58\pm0.27$ & 156 & 2 & 58640 \\

BPS BS 17569-049 & $4794\pm6$ & 0.80 & $-2.85\pm0.07$ & $2.65\pm0.24$ & 75 & 4 & 40970 \\

BPS CS 22953-003 & $5160\pm12$ & 2.35[+0.05] & $-2.63\pm0.06$ & $1.46\pm0.15$ & 69 & 3 & 40970 \\

HD 186478 & $4793\pm22$ & 1.76[$-0.26$] & $-2.27\pm0.09$ & $1.56\pm0.10$ & 141 & 2 & 40970 \\

HD 45282 & $5321\pm23$ & 3.22 & $-1.43\pm0.08$ & $1.18\pm0.06$ & 384 & 2 & 40970 \\

BPS CS 22892-052 ($\spadesuit$) & $4979\pm38$ & 1.68[+0.19] & $-2.86\pm0.08$ & $1.70\pm0.08$ & 107 & 2 & 40970 \\

\hline
\end{tabular}
\begin{tablenotes}
\item{} \textbf{Notes.} {
$\spadesuit$ indicates a CEMP star \citep{Sneden94}, not used in the analysis or plots.
Uncertainties in \teff\ are internal.
The third column lists surface gravities derived from non-LTE ionization equilibrium; values in brackets indicate the differences relative to those obtained from the fundamental formula in Sect.~\ref{sec:logg}.
The column headed ``N'' indicates the number of spectra co-added.
The last column indicates the spectral resolving power.
}
\end{tablenotes}
\end{threeparttable}
\end{table*}

\begin{table*}
\caption{Eu and Th abundances and abundance ratios for our sample stars.
The Th-to-Eu abundance ratios are calculated directly from the
reported absolute abundances.}
\label{tab:Eu_Th}
\centering
\tiny
\begin{threeparttable}
\begin{tabular}{l|ccccccc}
\hline\hline
Star & $A(\mathrm{Eu})_{\rm LTE}$ & $A(\mathrm{Eu})_{\rm NLTE}$ &
$A(\mathrm{Th})$ &
$\log(\mathrm{Th/Eu})_{\rm LTE}$ &
$\log(\mathrm{Th/Eu})_{\rm NLTE}$ &
[Eu/Fe] & [Th/Fe] \\
 & [dex] & [dex] & [dex] & [dex] & [dex] & [dex] & [dex] \\
\hline

BD+09 2860
& $-0.63^{+0.06}_{-0.09}$ [$\pm0.022$]
& $-0.61^{+0.06}_{-0.09}$
& $-1.08^{+0.08}_{-0.15}$
& $-0.45^{+0.09}_{-0.16}$
& $-0.47^{+0.09}_{-0.16}$
& $+0.44^{+0.09}_{-0.11}$
& $+0.47^{+0.11}_{-0.17}$ \\

BD+11 2998
& $-0.34\pm0.04$  [$\pm0.051$]
& $-0.27\pm0.04$
& $-0.74\pm0.09$
& $-0.40^{+0.10}_{-0.13}$
& $-0.47^{+0.10}_{-0.13}$
& $+0.24\pm0.10$
& $+0.27\pm0.13$ \\

BD+17 3248
& $-0.48^{+0.01}_{-0.02}$  [$\pm0.022$]
& $-0.44^{+0.01}_{-0.02}$
& $-0.96^{+0.07}_{-0.08}$
& $-0.48^{+0.07}_{-0.09}$
& $-0.52^{+0.07}_{-0.09}$
& $+0.91^{+0.07}_{-0.07}$
& $+0.89^{+0.10}_{-0.11}$ \\

BPS BS 17569-049
& $-1.65^{+0.06}_{-0.01}$  [$\pm0.014$]
& $-1.57^{+0.06}_{-0.01}$
& $-1.99^{+0.18}_{-0.37}$
& $-0.34^{+0.18}_{-0.37}$
& $-0.42^{+0.18}_{-0.37}$
& $+0.76^{+0.09}_{-0.07}$
& $+0.84^{+0.19}_{-0.38}$ \\

BPS CS 22953-003
& $-1.16\pm0.01$  [$\pm0.018$]
& $-1.10\pm0.01$
& $-1.47^{+0.30}_{-\infty}$
& $-0.31^{+0.30}_{-\infty}$
& $-0.37^{+0.30}_{-\infty}$
& $+1.01\pm0.06$
& $+1.14^{+0.31}_{-\infty}$ \\

HD 186478
& $-1.29\pm0.01$  [$\pm0.027$]
& $-1.24\pm0.02$
& $-1.82^{+0.10}_{-0.21}$
& $-0.53^{+0.10}_{-0.21}$
& $-0.58^{+0.10}_{-0.21}$
& $+0.51\pm0.09$
& $+0.43^{+0.13}_{-0.23}$ \\

HD 45282
& $-0.40^{+0.01}_{-0.07}$  [$\pm0.027$]
& ---
& $-0.88^{+0.06}_{-0.05}$
& $-0.48^{+0.06}_{-0.05}$
& ---
& $+0.51^{+0.08}_{-0.11}$
& $+0.53^{+0.10}_{-0.09}$ \\

BPS CS 22892-052
& $-0.78\pm0.01$  [$\pm0.041$]
& ---
& $-1.41^{+0.16}_{-0.19}$
& $-0.63^{+0.20}_{-0.17}$
& ---
& $+1.56\pm0.08$
& $+1.43^{+0.18}_{-0.21}$ \\

\hline
\end{tabular}
\begin{tablenotes}
\item[] \textbf{Notes.}
{
For A(Eu)$_{\mathrm{LTE}}$, the uncertainties arising from
\teff\ and \logg\ are reported in parentheses and should
be added in quadrature to the quoted abundance
uncertainties (line-to-line dispersions) to obtain the
total uncertainties. For A(Th), the contributions from
\teff\ and \logg\ are negligible compared to the
uncertainty associated with line fitting.
The abundance ratios [Eu/Fe] and [Th/Fe] were calculated
adopting solar abundances of
$A(\mathrm{Eu})_\odot=0.52$ and
$A(\mathrm{Th})_\odot=0.02$
from \citet{asplund2009ARA&A..47..481A}.
Their uncertainties were obtained by adding in
quadrature the corresponding uncertainties in A(Eu)
and A(Th) with that in [Fe/H].
}
\end{tablenotes}
\end{threeparttable}
\end{table*}

\section{Testing the Production Ratio}
\label{sec:PR}

The PR is defined as the initial abundance ratio  of two elements produced, e.g. by the r-process, before any radioactive decay has taken place. In the case of Th and Eu, in a logarithmic scale it can be expressed as
${\rm PR}= \log(\mathrm{Th}/\mathrm{Eu})_{\mathrm{PR}}$.

The PR represents the nucleosynthetic initial condition of the Th/Eu chronometer and is therefore the key reference point for age determinations based on radioactive decay. Since thorium is a radioactive element while europium is stable, the observed abundance ratio between the two elements evolves with time according to
\begin{equation}
\log(\mathrm{Th}/\mathrm{Eu})_{\mathrm{obs}} = PR - \frac{t}{\tau \ln 10},
\label{eq:ThEu_Age}
\end{equation}
 where t is time and $\tau = 20.3 \pm 0.1$~Gyr is the mean lifetime
($\tau = T_{1/2}/\ln 2$) of $^{232}$Th, corresponding to a half-life
of $T_{1/2}=14.05\pm0.06$~Gyr \citep[e.g.][]{Cowan1999ApJ...521..194C}.

At low metallicity ($\mathrm{[Fe/H]} \lesssim -1.5$~dex), the PR is commonly assumed to be approximately constant \citep[see][for a direct estimate of the PR at different metallicities]{Azhari2025A&A...699A.276A}. In this regime, the interstellar medium was enriched by a relatively small number of r-process events, and Galactic chemical evolution had not yet averaged over many nucleosynthetic sources. 
This assumption provides the basis for using Th/Eu as a chronometer, but can be tested observationally by examining the dispersion of decay-corrected ratios.

If the main r-process operates robustly, individual enrichment events are expected to produce similar relative abundance patterns for the heavy r-process elements. Consequently, the resulting abundance pattern can be approximated by a single representative PR for early Galactic material \citep{Sneden2008ARA&A..46..241S,frebel2018ARNPS..68..237F}, although intrinsic event-to-event variations cannot be excluded, particularly for the actinides (e.g., actinide-enhanced and actinide-deficient stars). 

In this work, we adopt an r-process PR of $\log(\mathrm{Th/Eu})_{\rm PR} = -0.24$, which is consistent with a broad range of theoretical r-process nucleosynthesis models and empirical constraints \citep{Cowan1999ApJ...521..194C,Kratz2007PrPNP..59..147K,Ji2018ApJ...856..138J, Bauswein2013ApJ...773...78B,  Azhari2025A&A...699A.276A}, and corresponds to the WP1 model of \citet{Hill2017A&A...607A..91H}. An uncertainty of $\pm0.10$ dex is adopted as a representative estimate of the theoretical spread in the predicted Th/Eu PR. 

The adopted value is close to the recent observational estimate (PR=$-0.27\pm0.11$) obtained in a homogeneous manner within the R-Process Alliance project\footnote{\url{https://sites.google.com/view/rprocessalliance/home?authuser=0}} in S26 with the largest sample of stars with Th and Eu abundance measurements, and validated for the different classes of RPE stars. Under the assumption of a constant PR at low metallicity, variations in the observed $\log(\mathrm{Th}/\mathrm{Eu})$ are interpreted primarily as the result of radioactive decay of Th rather than intrinsic differences in the initial production conditions, making the PR the fundamental anchor of $\log({\rm Th}/{\rm Eu})$ as a nucleocosmochronometer.

\subsection{The PR in our sample}

An important test of this framework is to investigate also in our sample  whether the observed dispersion of $\log(\mathrm{Th}/\mathrm{Eu})$ ratios is consistent with a common representative PR within the uncertainties.  We restrict the analysis to log($\mathrm{Th}/\mathrm{Eu}$) in stars that are not affected by known abundance anomalies (actinide boost and deficient). In particular, as already mentioned, we excluded CEMP stars, particularly those whose chemical abundances have been altered by mass transfer from a binary companion.

By selecting metal-poor stars without anomalies in the actinide-to-lanthanide ratios, regardless of their overall r-process enrichment level, we can test whether the observed scatter in $\log(\mathrm{Th}/\mathrm{Eu})$ at low metallicity is consistent with measurement uncertainties and radioactive decay alone. 
The assumption of a common representative PR can be tested by examining the residual dispersion of $\log(\mathrm{Th}/\mathrm{Eu})$ after accounting for radioactive decay. 

\begin{figure*}
\sidecaption
    \centering
    \includegraphics[width=12cm]{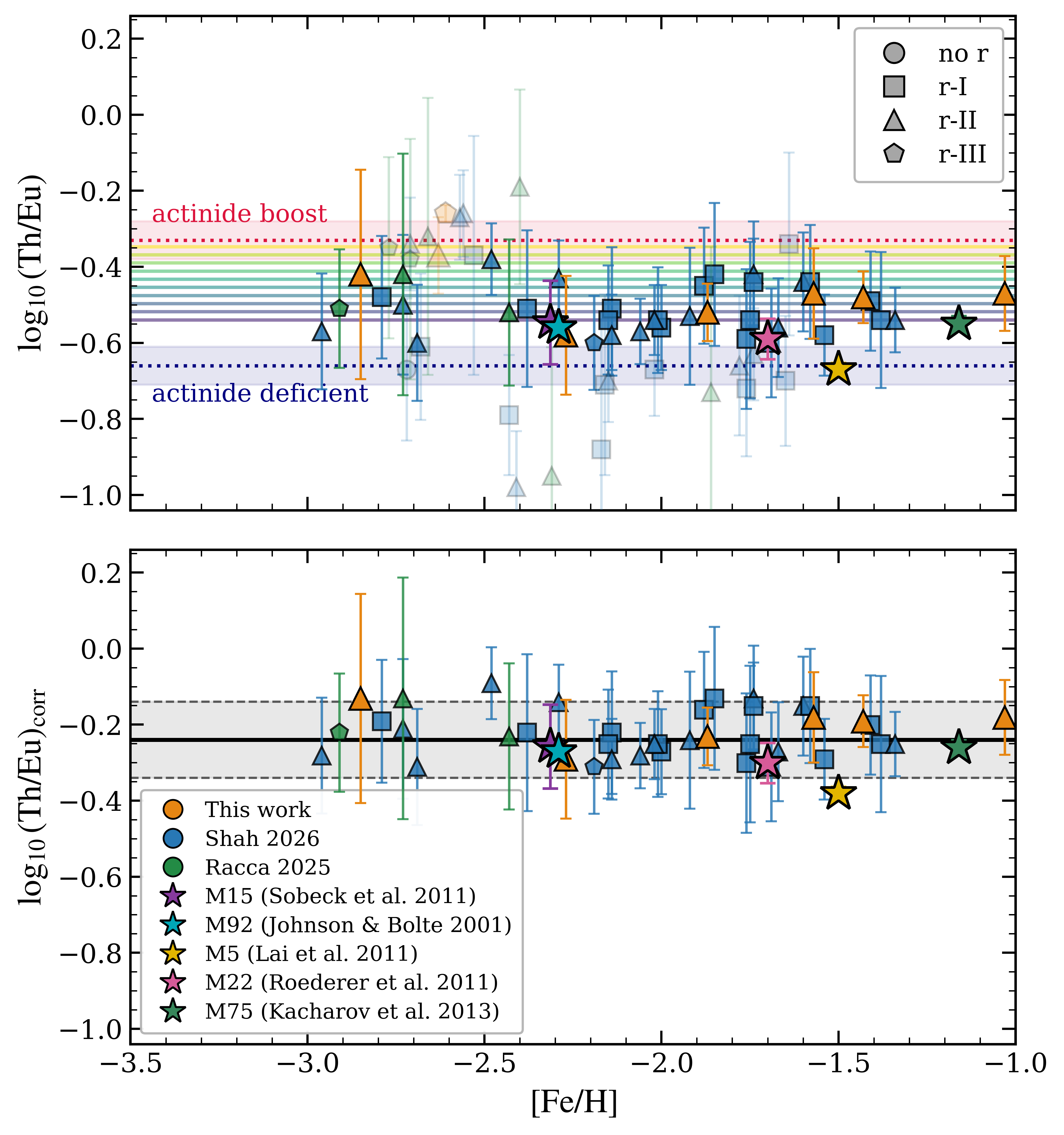}
   \caption{
Top panel: observed $\log({\rm Th}/{\rm Eu})$ abundance ratios for our stellar sample (orange symbols) combined with the homogeneous metal-poor sample of \citet{Racca2025A&A...704A.282R} (green symbols) and of \citet{Shah2026arXiv260412892S} (blue symbols).    
Shaded horizontal bands show the thresholds used to identify actinide-boost and actinide-deficient stars ($\log(\rm{Th/Eu}) = -0.33 \pm 0.05$ and $-0.66 \pm 0.05$, respectively). Coloured curves represent theoretical radioactive decay tracks corresponding to ages from 5 to 14 Gyr (from top to bottom).  
Bottom panel: $\log({\rm Th}/{\rm Eu})$ ratios corrected for radioactive decay assuming an age of 13.5 Gyr. The horizontal solid line marks the theoretical PR $=-0.24 \pm 0.10$, while dashed lines indicate the associated uncertainty. In both panels, different symbols indicate r-process classification subclasses.
}
    \label{fig:theu_evolution}
\end{figure*}

To do this, we directly plot in the upper panel of Fig.~\ref{fig:theu_evolution} the observed $\log({\rm Th}/{\rm Eu})$  abundance ratios of our stellar sample, combined with the homogeneous samples of metal-poor stars from R25 and S26, both from the R-process alliance project.  We also include in the figure the globular clusters for which Th measurements are available: M15 from \citet{Sobeck2011AJ....141..175S}, M92 from \citet{Johnson2001ApJ...554..888J}, M5 from \citet{Lai2011AJ....141...62L},  M22 from \citet{Roederer2011ApJ...742...37R} and M75 from \citet{Kacharov2013A&A...554A..81K}.
Additional upper limits on the Th/Eu ratio have been reported for M2 and NGC 4833 by \citet{Roederer2015MNRAS.449.3889R}, and for NGC 5824 by \citet{roederer2016ApJ...821...37R}, which are not shown in the the figure.

We adopt the same thresholds used by S26 to identify actinide-boost and actinide-deficient stars, namely $\log(\rm{Th/Eu}) = -0.33 \pm 0.05$ and $-0.66 \pm 0.05$, respectively. Following the above definition of anomalous stars, about 40\% of the combined sample (the present sample, R25 and S26) is excluded from the comparison with the theoretical PR. 
We also include theoretical decay tracks anchored to the empirical PR. 
Our combined sample spans a metallicity range approximately from [Fe/H] $\sim -3$ to $-1.3$~dex.
By comparing the observed abundances with the decay curves corresponding to ages between 5 and 14 Gyr (shown from top to bottom, one line per Gyr in Fig.~\ref{fig:theu_evolution}), we find that the sample is broadly consistent with this age range, with most stars (approximately 65\%) consistent with the oldest ages considered.

In the bottom panel, we correct the $\log({\rm Th}/{\rm Eu})$ abundance ratios for the radioactive decay of thorium, assuming representative stellar ages of 13.5 Gyr. We find negligible differences when adopting instead the approach of S26, who assumed ages of 13.2 Gyr for [Fe/H] $< -2.2$, 13.0 Gyr for $-2.2 <$ [Fe/H] $< -1.5$, and 12 Gyr for [Fe/H] $> -1.5$, or a constant average age of 13 Gyr.
 The corrected abundance ratios show excellent agreement with the theoretical PR value for all three samples: the S26 and R25  samples and our sample. 
This exercise provides a consistency check of the adopted theoretical PR rather than an independent empirical calibration, as the decay correction depends on the assumed stellar age. The inferred PR shows no significant dependence on metallicity or r-process classification (no r, r-I, r-II, or r-III), consistent with S26 and supporting the use of $\log({\rm Th}/{\rm Eu})$ as a nucleocosmochronological age indicator.

\section{Ages from nucleocosmochronology}
\label{sec:ages}

In this Section, we compute nucleocosmochronological ages for our sample and 
the recent literature  samples of \citet{Racca2025A&A...704A.282R} and \citet{Shah2026arXiv260412892S}. We investigate the limitations introduced by moderate signal-to-noise ratios and spectra with  R$\sim$40,000, evaluate the impact of uncertainties in $\log({\rm Th}/{\rm Eu})$ on the derived ages, and discuss possible improvements for future nucleocosmochronological applications.

\subsection{Ages for our sample and the literature samples}

In the previous section, we adopted a representative age for metal-poor stellar populations to estimate an empirical initial PR. The resulting value was found to be consistent, within the uncertainties, with theoretical predictions. We stress that this procedure is intended as a consistency test of the adopted theoretical PR rather than as an independent empirical calibration, since the correction for radioactive decay itself requires an assumption about stellar age. For the age determination, we therefore adopt the theoretical value $\log({\rm Th}/{\rm Eu})_{\rm PR}=-0.24$ discussed in Sect.~\ref{sec:PR} and use it to derive individual nucleocosmochronological ages for both our sample and the literature stars.

The uncertainty in the adopted PR is not included in the quoted individual age uncertainties, as it represents a systematic uncertainty in the absolute zero-point of the age scale and does not affect relative age differences among stars analysed using the same PR. We instead propagate the uncertainties in the measured $\log({\rm Th}/{\rm Eu})$ abundance ratios, accounting for the observational contributions to the Th and Eu abundance errors (see also \citet{Ludwig2010A&A...509A..84L} for details on the error budget). The quoted age uncertainties should therefore be interpreted as observational uncertainties rather than as total uncertainties on the absolute age scale, which is additionally affected by the systematic uncertainty in the adopted PR.

\begin{table}[t]
\centering
\caption{Stellar ages derived from the Th/Eu chronometer, adopting an initial production ratio of $\log(\mathrm{Th/Eu})_0=-0.24$. Ages are based on NLTE Eu abundances where available, while LTE abundances are used for HD 45282. Uncertainties are propagated from the measured abundance ratios and do not include uncertainties in the production ratio.}
\label{tab:ThEu_ages}
\small
\begin{tabular}{lc}
\hline\hline
Star & Age [Gyr] \\
\hline
PS BS 17569-049 & $8.4 \pm 12.8$ \\
BD+11 2998 & $10.7 \pm 4.6$ \\
BD+09 2860 & $10.7 \pm 5.6$ \\
HD 45282 & $11.2 \pm 3.2$ \\
BD+17 3248 & $13.1 \pm 3.5$ \\
HD 186478 & $15.9 \pm 7.3$ \\
\hline
\end{tabular}
\end{table}

\begin{figure}
    \centering
    \includegraphics[width=0.99\linewidth]{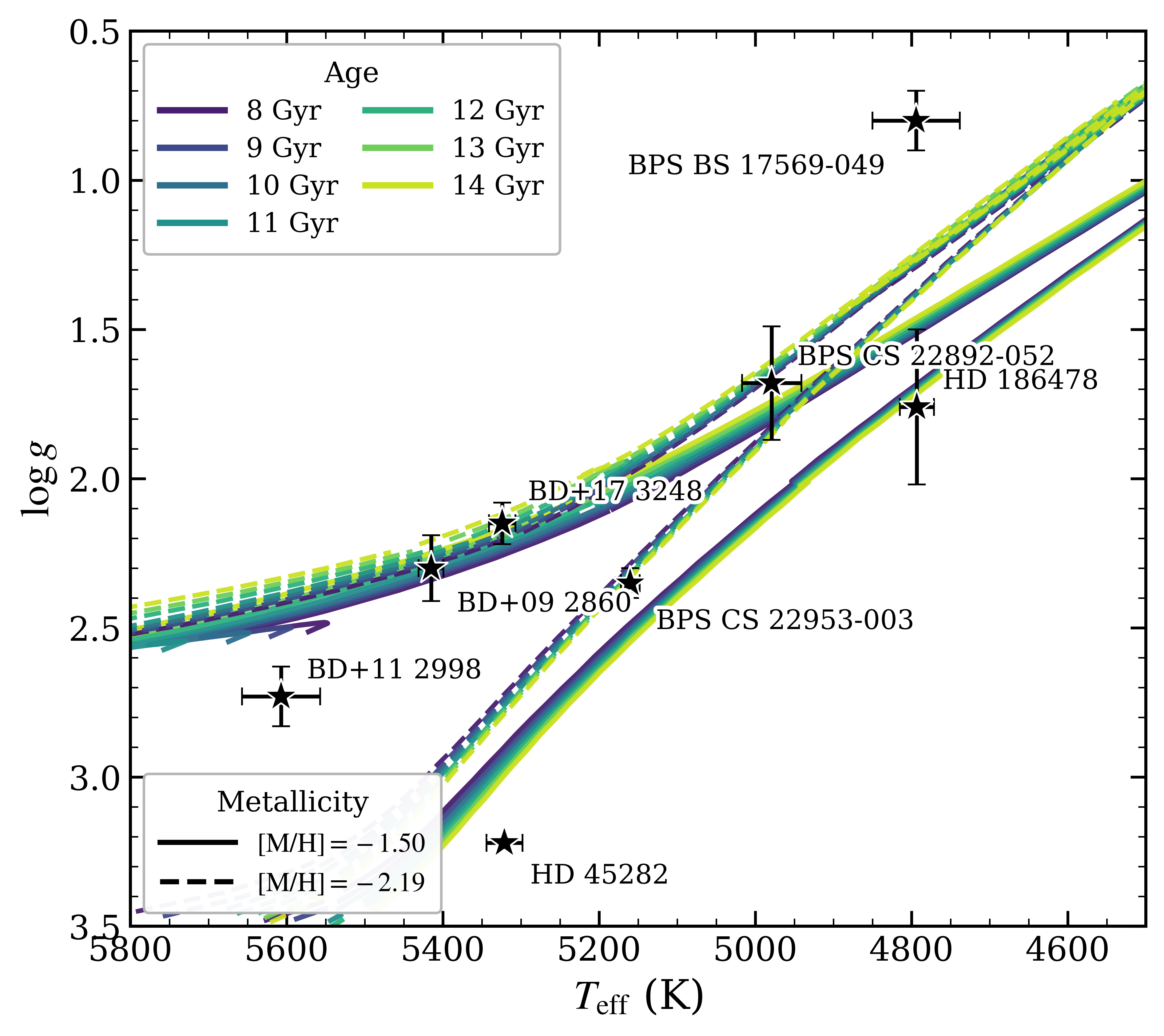}
    \caption{Kiel diagram showing the stars in our sample (black crosses) compared with {\sc parsec} 3.9 isochrones. For each metallicity, the seven isochrones correspond to ages of 8, 9, 10, 11, 12, 13, and 14 Gyr. Solid and dashed lines indicate [M/H] = $-1.5$ and $-2.19$~dex, respectively.}
    \label{fig:isochrones}
\end{figure}

The ages of the stars in our sample, excluding the CEMP-r star BPS CS 22892-052 and the actinide-boost star BPS CS 22953-003, are presented in Table~\ref{tab:ThEu_ages}. In Fig.~\ref{fig:isochrones}, we show the location of all stars in our sample in the Kiel diagram, together with eight isochrones from {\sc parsec} 3.9 \citep{Bressan2012MNRAS.427..127B}. We consider two metallicities, [M/H] = $-1.5$ and $-2.19$~dex, the latter corresponding to the lower metallicity limit of the {\sc parsec} isochrones. For each metallicity, we consider seven ages from 8 to 14 Gyr in steps of 1 Gyr, representative of the age and metallicity ranges covered by our sample. The isochrones are highly degenerate along the RGB, with differences between the 11 and 14 Gyr tracks that are too small to provide reliable individual age constraints for our stars. The comparison therefore only allows us to conclude that the stars in our sample are consistent with an old stellar population.

\begin{figure}
    \centering
    \includegraphics[width=0.99\linewidth]{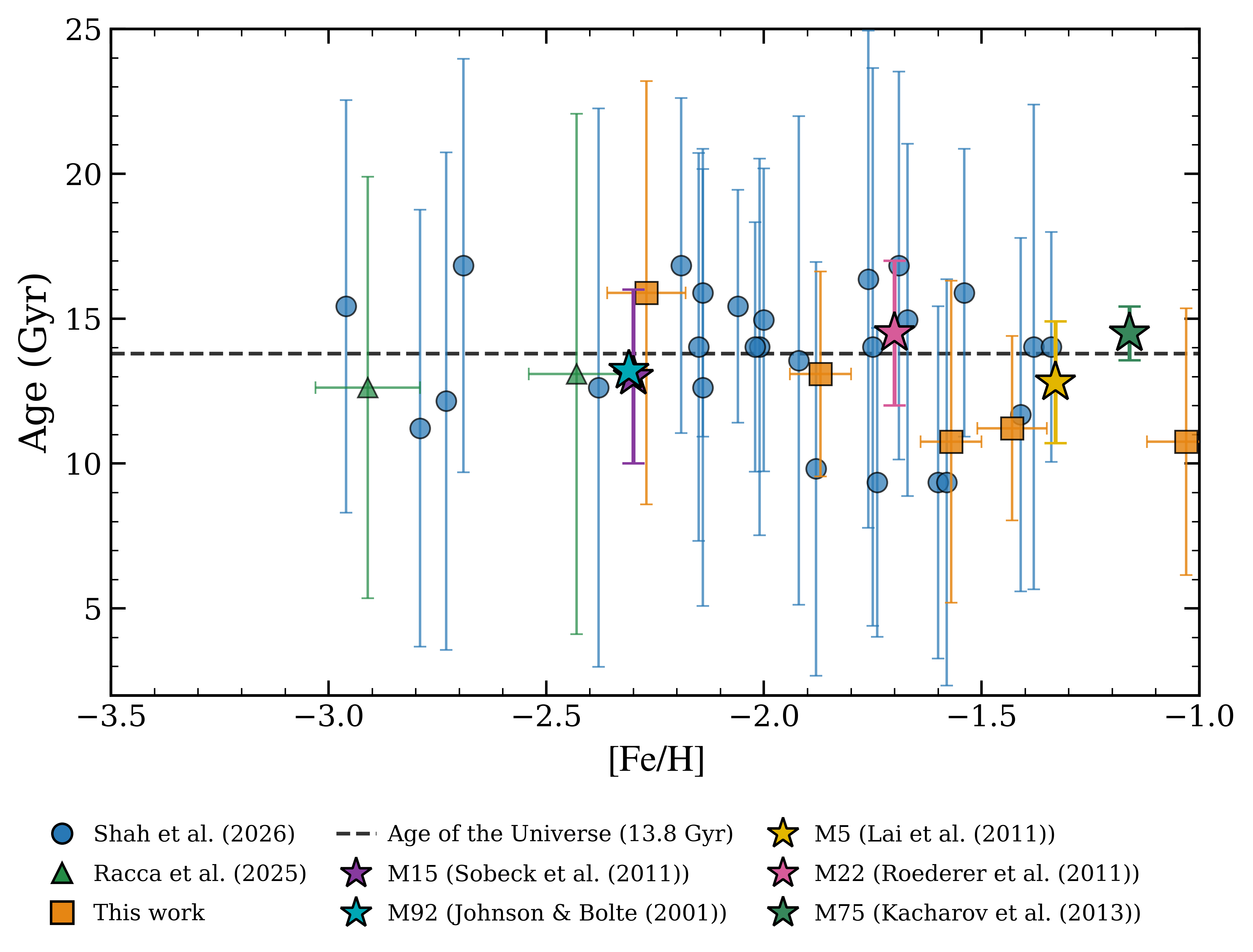}
    \caption{Age–metallicity relation for our sample and for the literature  samples, computed assuming the radioactive decay of Th abundances. The horizontal black dashed line indicates the assumed age of the Universe (13.8 Gyr). In blue the individual ages for the S26 sample, in green for the R25 one, and in orange those from the present work.
   Actinide-boost and actinide-deficient stars were excluded, and the comparison was restricted to stars with inferred nucleocosmochronological ages older than 8 Gyr.  The globular clusters with Th measurements are indicated by star symbols. For clusters with multiple Th measurements, the error bars correspond to the standard deviation of the mean age derived from the individual member stars; no error bar is shown when only a single star is available. }
    \label{fig:ages}
\end{figure}

The nucleocosmochronological ages of our sample and the literature stars are shown in Fig.~\ref{fig:ages} as a function of [Fe/H]. The 
resolution of the spectra used for the Th abundance measurements ($R \simeq 40,000$--$50,000$), combined with their moderate signal-to-noise ratios for some stars, results in relatively large uncertainties in the derived $\log({\rm Th}/{\rm Eu})$ ratios and, consequently, in the individual nucleocosmochronological ages, typically amounting to several Gyr.

Since our aim is to investigate the oldest stellar populations, we restrict the following comparison to stars with inferred nucleocosmochronological ages older than 8 Gyr. The resulting subsample contains 33 stars and has a median age of  $14.02^{+1.87}_{-3.22}$ Gyr. This value is consistent, within the quoted observational uncertainties, with the age of the Universe, $13.797\pm0.023$~Gyr, predicted by the standard $\Lambda$CDM cosmological model \citep{Planck2020A&A...641A...6P}. We emphasise, however, that this median characterises the selected old subsample and should not be interpreted as an independent determination of the age of the oldest stellar populations or of the Universe. Moreover, the systematic uncertainty associated with the adopted PR is not included in the quoted uncertainty and affects the absolute zero-point of the nucleocosmochronological age scale.

A similar behaviour is observed for the globular clusters with available Th abundances: M15 \citep{Sobeck2011AJ....141..175S}, M92 \citep{Johnson2001ApJ...554..888J}, M5 \citep{Lai2011AJ....141...62L},  M22 \citep{Roederer2011ApJ...742...37R} and M75 \citep{Kacharov2013A&A...554A..81K}. Their nucleocosmochronological ages are, within the uncertainties, consistent with those expected for ancient stellar populations.

No clear correlation between age and metallicity is observed in the combined sample within the current uncertainties. This is consistent with recent studies showing a substantial age dispersion at fixed metallicity \citep{Bonifacio2026arXiv260617674B}. In particular, their analysis indicates that, for ages older than $\sim8$ Gyr, the age--metallicity relation becomes highly dispersed, with metallicity providing only a weak constraint on stellar age.

\subsection{Estimating the sensitivity of $\log({\rm Th}/{\rm Eu})$  for ages}

\begin{figure*}
    \sidecaption
    \centering
    \includegraphics[width=12cm]{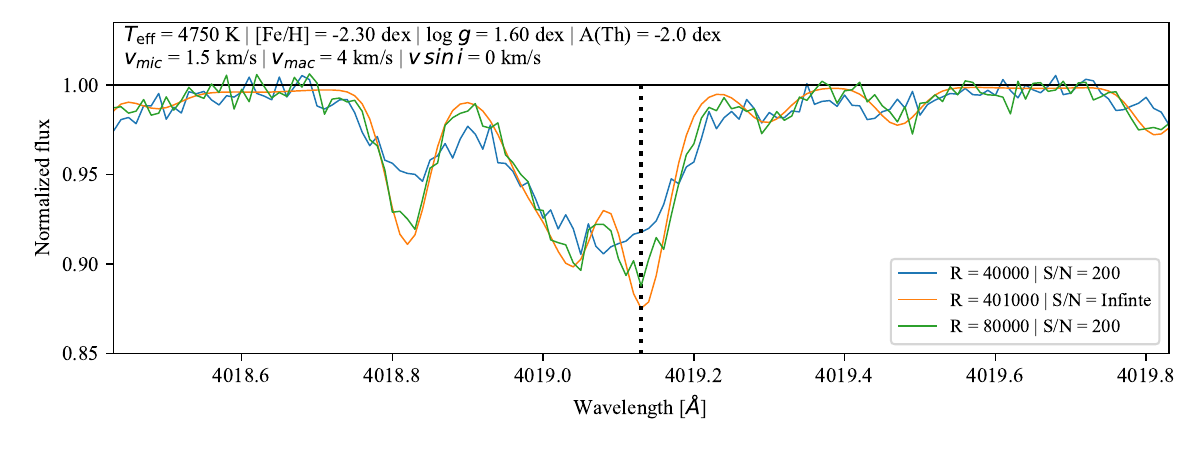}
    \caption{Spectra simulated for HD~186478. The  atmospheric parameters assumed in the synthesis are indicated in the plot. Ni, Ce, and Co abundances adopted are solar-scaled. All spectra were convoluted with $v_{\mathrm{mac}} = 4$ km s$^{-1}$, which was retrieved from the real spectrum of HD~186478 in the fit of Fig~\ref{fig:plots}. } The Th~{\sc ii} 4019~\AA\ line is indicated by the vertical dotted line. Different combinations of spectral resolution and S/N ratio are reported in the legend.  The noise-free spectrum at $R=401,000$ is shown as a reference.
    \label{fig:synthesisMP}
\end{figure*}

\begin{figure}
    \centering
    \includegraphics[width=1.0\linewidth]{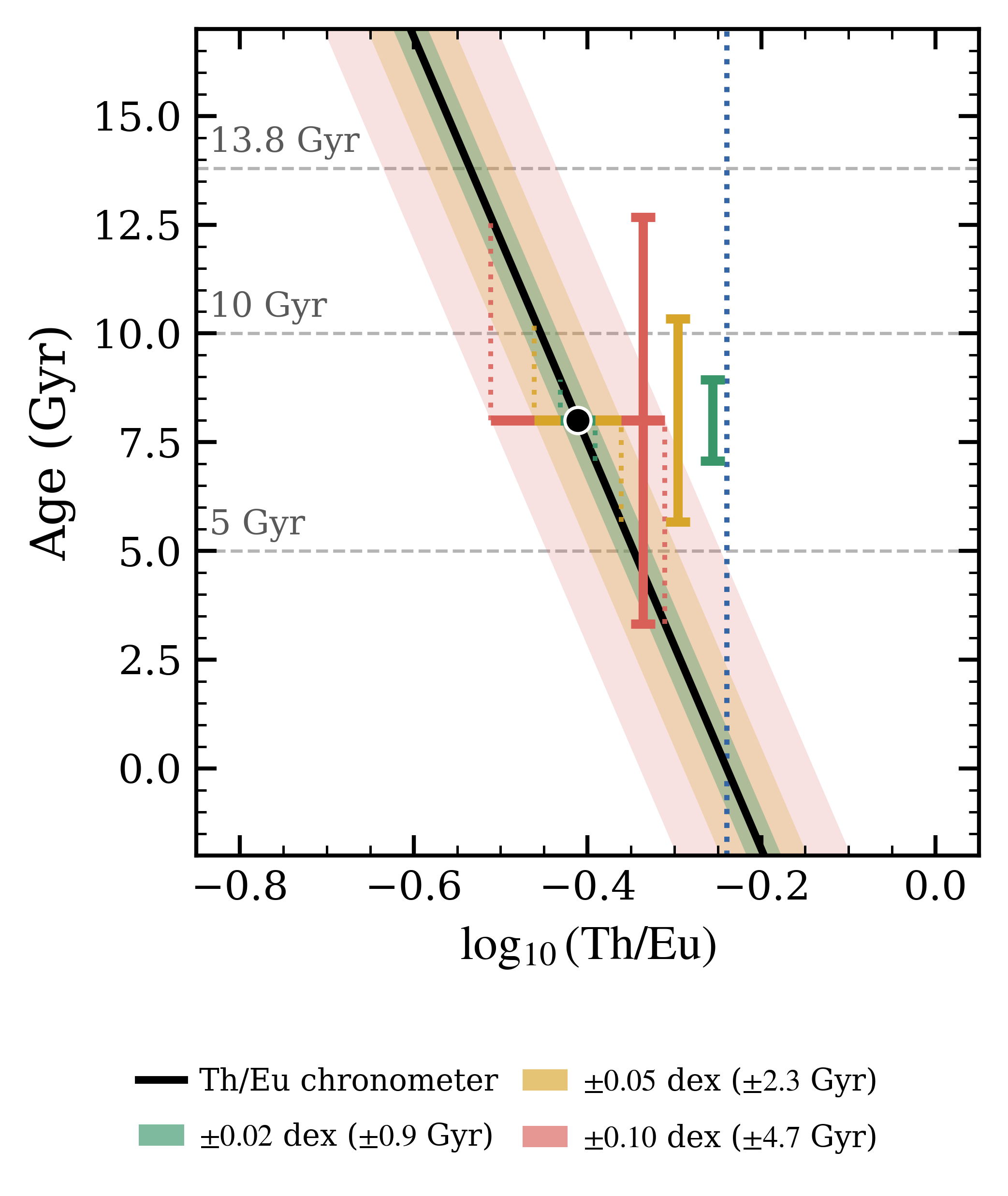}
    \caption{Sensitivity of the $\log({\rm Th}/{\rm Eu})$  chronometer to stellar ages. The relation between the logarithmic abundance ratio $\log({\rm Th}/{\rm Eu})$  and the inferred age is shown assuming a fixed PR. The slope reflects the radioactive decay of 
$^{232}$Th, with a characteristic timescale set by $\tau$ln10. Shaded regions illustrate the impact of typical observational uncertainties in $\log({\rm Th}/{\rm Eu})$, translated into age uncertainties. The black dot illustrates how uncertainties in $\log({\rm Th}/{\rm Eu})$ translate into age uncertainties for a star with an age of 8 Gyr. The vertical dotted blue line indicates the PR. }
    \label{fig:sensitivity}
\end{figure}

Assuming that the PR is known and constant, nucleochronological ages depend 
critically on the accuracy and precision of the Th and Eu abundance measurements. 
The impact of observational uncertainties on the precision of 
nucleochronological ages has been extensively discussed by 
\citet{Ludwig2010A&A...509A..84L}. 
In particular, the Th line is prone to blending with nearby atomic and molecular 
features, and the precision of the derived thorium abundance strongly depends on 
the spectral resolution and quality of the observed spectra. Increasing the resolution from high-resolution ($R\simeq40\,000$) to very high-resolution spectroscopy is particularly important for stars without extreme Th enhancement. In Fig.~\ref{fig:synthesisMP}, we show synthetic spectra at different resolutions for a star with atmospheric parameters and abundances similar to HD~186478, focusing on the spectral region around the Th~{\sc ii} line at 4019~\AA.
We adopted HD~186478 as representative, although to assert this a larger stellar sample would be required. Nevertheless, except for the CEMP star, the spectra in Fig.~\ref{fig:plots} show similar similar contributions from Th to the blended feature. More generally, the measurability of Th in this spectral region depends primarily on the relative contributions of the neighbouring Fe and $^{13}$CH features: the stronger these contributions, the more difficult it becomes to identify and reliably measure the Th abundance.

Increasing the spectral resolution from $R\sim40,000$ to $R=80,000$ significantly improves the precision of Th abundance measurements by reducing blending around the Th~{\sc ii} 4019~\AA\ feature \citep[see, e.g.][]{Kordopatis2023A&A...674A.104K}. Our simulations (Appendix~\ref{sec:th_profiles_err}) show that the typical uncertainty in $\log({\rm Th}/{\rm Eu})$ decreases from $\sim0.10$--$0.15$ dex to $\sim0.04$--$0.05$ dex (Fig.~\ref{fig:err_th}), reducing the corresponding age uncertainty from $\sim5$--$7$ Gyr to $\sim2$--$3$ Gyr (Fig.~\ref{fig:sensitivity}). This improvement persists even at a fixed photon budget, despite the lower S/N per resolution element at higher resolution (Fig.~\ref{fig:synthesisMP}), demonstrating that increasing spectral resolution is more effective than increasing S/N alone.

As shown in Fig.~\ref{fig:sensitivity}, achieving an age precision of approximately 2.5 Gyr requires a combined uncertainty in $\log({\rm Th}/{\rm Eu})$ of about 0.05 dex. This implies that the uncertainty in $A({\rm Th})$ should be limited to approximately 0.04 dex, assuming a comparable or smaller contribution from the Eu abundance. The gain provided by higher spectral resolution is therefore particularly relevant for nucleocosmochronological applications, where the weak and blended Th~{\sc ii} features currently represent one of the main observational limitations.

\subsection{Improving nucleocosmochronological ages in coeval stellar populations}

The precision of nucleocosmochronological ages can be further improved
when applied to coeval stellar populations, such as globular clusters,
as already discussed by \citet{Ludwig2010A&A...509A..84L}.
Globular clusters are particularly well suited for this purpose, as
their stars are expected to share a common formation epoch and to have
experienced a broadly similar early chemical enrichment
history\footnote{The age differences between multiple populations in
globular clusters are generally small compared to the systematic
uncertainties affecting isochrone-based absolute ages, allowing
globular clusters to be treated as effectively coeval systems for
nucleocosmochronological applications
\citep[e.g.][]{VandenBerg2013ApJ...775..134V}.}.
Their ages can also be independently constrained through isochrone
fitting, with homogeneous studies indicating that most Galactic
globular clusters belong to an old stellar population with ages of
$\sim$12--13 Gyr
\citep[e.g.][]{VandenBerg2013ApJ...775..134V,Massari2023A&A...680A..20M},
or through other chemical tracers, such as beryllium
\citep{Pasquini2004A&A...426..651P}.

Individual nucleocosmochronological measurements within a cluster can
therefore be regarded as multiple estimates of the same underlying
formation epoch rather than as independent stellar ages.
This coevality provides both a statistical advantage and a direct
empirical test of the internal consistency of the Th/Eu chronometer.
The observed star-to-star dispersion in $\log({\rm Th}/{\rm Eu})$ in a
globular cluster can be compared with that expected from the individual
measurement uncertainties.
If the observed scatter is consistent with these uncertainties and is
dominated by random errors, combining multiple stars reduces the
statistical uncertainty on the mean approximately as $1/\sqrt{N}$,
where $N$ is the number of stars analysed, allowing a more precise
determination of the mean nucleocosmochronological age.
Systematic uncertainties shared by all cluster members, however,
do not decrease with increasing sample size and therefore set a floor
to the precision achievable through statistical averaging alone.

Conversely, an excess star-to-star scatter beyond that expected from
the individual uncertainties would point to star-dependent systematic
effects or possible intrinsic variations in the r-process abundance
pattern.
Moreover, when the cluster age is independently constrained, for
example through isochrone fitting, the mean
$\log({\rm Th}/{\rm Eu})$ ratio and its star-to-star distribution can
provide constraints on theoretically predicted PRs.

To date, however, the potential of this approach has been explored
only for a limited number of stars.
No clear actinide-boost signature has been identified among globular
cluster stars; however, thorium has been detected in only a few
clusters, including M5, M15, M22, M75, and M92
\citep{Sneden2000ApJ...536L..85S,Johnson2001ApJ...554..888J,
Lai2011AJ....141...62L,Sobeck2011AJ....141..175S,
Roederer2011ApJ...742...37R,Kacharov2013A&A...554A..81K,
Roederer2015MNRAS.449.3889R}.
Where available, ages inferred from measured $\log({\rm Th}/{\rm Eu})$ ratios are generally consistent, within their substantial uncertainties, with independent age estimates obtained from stellar evolution models, as shown for M15 by \citet{Sneden2000ApJ...536L..85S} and for a sample of globular clusters by \citet{Roederer2015MNRAS.449.3889R}, who compared Th/Eu-based ages with the isochrone ages of \citet{Mar2009ApJ...694.1498M}.
These results support the use of globular clusters as benchmark
systems for radioactive chronometers, but the small number of stars
with reliable Th abundances currently limits both the statistical gain
offered by coeval populations and the possibility of characterising
their intrinsic star-to-star dispersion.

Increasing the sample size alone, however, cannot reduce all
components of the error budget.
New proposed instruments, such as the high-resolution multi-object spectrograph  {\sc hrmos}
\citep{magrini2023arXiv231208270M,magrini26,
Bianco2026arXiv260722411B,Randich2026vltb.confE..44R,
Riva2026vltb.confE..54R},
would provide a complementary advantage by improving the precision
of individual abundance measurements while enabling observations of
large stellar samples.
Its combination of high throughput in the blue spectral region,
multiplex capability, and a resolving power of $R=80,000$ would
enable high-S/N measurements of Th and Eu for hundreds of stars
within individual globular clusters.
The improved spectral quality would reduce the observational
uncertainties of individual abundance measurements and allow a more
accurate treatment of line blending, continuum placement, and
spectral-line fitting, all of which are particularly important for
the weak Th features.
Once this observational component of the error budget is reduced,
large samples of coeval stars would also make it possible to place
tighter constraints on residual systematic effects, for example
those associated with atomic data, stellar-atmosphere models, or
line-formation modelling.

The combination of these two complementary advantages---the
statistical leverage provided by coeval stellar populations and
the improved precision of individual measurements enabled by
{\sc HRMOS}---would make globular clusters powerful laboratories
for Th/Eu nucleocosmochronology.
Measurements of hundreds of stars per cluster could simultaneously
provide more robust mean ages, test whether the observed star-to-star
scatter is consistent with the estimated uncertainties, and search
for intrinsic variations in the r-process abundance pattern.
Such observations would therefore provide stringent empirical tests
of the assumptions underlying the Th/Eu chronometer and could
ultimately yield independent constraints on the formation times
of the oldest Galactic stellar populations.

The relevance of such measurements extends beyond the calibration
of the $\log({\rm Th}/{\rm Eu})$ chronometer.
The oldest globular clusters have historically provided independent
lower limits on the age of the Universe through stellar evolution
models
\citep{Krauss2003Sci...299...65K,Valcin2020JCAP...12..002V}.
Nucleocosmochronology provides an independent age diagnostic based
on radioactive decay rather than stellar evolution.
Applying both methods to the same systems therefore offers a
complementary way to constrain the ages of the oldest stellar
populations and the early formation history of the Milky Way,
while providing an independent consistency check between different
stellar dating techniques.

A major source of systematic uncertainty, however, is the
initial Th/Eu PR, whose theoretical value remains
uncertain.
For example, the neutron-star merger calculations of
\citet{Bauswein2013ApJ...773...78B} predict initial
$\log({\rm Th}/{\rm Eu})$ ratios ranging from approximately
$-0.27$ to $-0.15$ dex across the models considered, illustrating
the sensitivity of the predicted PR to the adopted equation of
state and associated ejecta conditions.
The full range corresponds to a difference of approximately
5.6 Gyr in the inferred absolute ages.
If a common initial PR is assumed for all populations, changing its adopted value produces a systematic shift in their absolute ages, however without affecting their relative ages.
Improved theoretical predictions of r-process nucleosynthesis
yields, particularly of the initial Th/Eu production ratio,
are therefore essential for further progress in absolute
nucleocosmochronology.

In combination with other age indicators, these measurements
could ultimately provide an independent lower bound on the age
of the Universe
\citep{Cowan1997ApJ...480..246C,Chaboyer1998PhR...307...23C,
Sneden2001Natur.409..673S,Krauss2003Sci...299...65K,
Valcin2020JCAP...12..002V,Cimatti2023ApJ...953..149C,
Tomasetti2026A&A...707A.111T,Verde2024ARA&A..62..287V}.

\section{Summary and conclusions}
\label{sec:summary}

In this work, we investigated the reliability of the $\log({\rm Th}/{\rm Eu})$ ratio as a nucleocosmochronometer for old stellar populations. By deriving new thorium and europium abundances for a sample of benchmark metal-poor stars from the \titans\ sample and combining them with recent homogeneous literature measurements, we explored the behaviour of Th relative to Eu, the validity of adopting a representative PR, and the potential of this chronometer to constrain the ages of the oldest stellar populations and provide an independent lower bound on the age of the Universe.

We derived Th and Eu abundances from high-resolution UVES spectra in eight stars using detailed spectral synthesis, accounting for the main atomic and molecular blends affecting the Th~{\sc ii} 4019~\AA\ feature. The resulting $\log({\rm Th}/{\rm Eu})$ ratios are consistent with previous homogeneous measurements and extend the available sample of metal-poor stars with reliable Th detections. Once stars with anomalous actinide abundances are excluded, the observed abundance ratios show no significant dependence on metallicity over the explored range ($-3 \lesssim {\rm [Fe/H]} \lesssim -1.3$). The agreement between our measurements and those from the R-Process Alliance samples further indicates that the relative production of Th and Eu is approximately stable across the metal-poor regime, within the current observational uncertainties and for stars without anomalous actinide abundances. Our analysis therefore supports the use of a common representative PR for these stars and, consequently, the application of $\log({\rm Th}/{\rm Eu})$ as an age indicator for old stellar populations.

We adopted the theoretical PR $\log({\rm Th}/{\rm Eu})_{\rm PR}=-0.24$ \citep{Hill2017A&A...607A..91H} and derived nucleocosmochronological ages for our sample and for literature stars. Although the uncertainties on individual ages remain large, typically several Gyr, mainly because of the limited precision of Th abundance measurements, the combined sample shows no clear age--metallicity trend within the current uncertainties, consistent with the behaviour found by \citet{Bonifacio2026arXiv260617674B} with ages obtained with different methods. These results illustrate that, even when individual stellar ages are poorly constrained, nucleocosmochronology can provide useful statistical information on the ages and formation history of ancient stellar populations.

Two complementary approaches emerge for improving the precision and applicability of the Th/Eu chronometer. First, increasing the resolving power from $R\sim40,000$ to $R\sim80,000$ substantially reduces the impact of line blending around the Th~{\sc ii} 4019~\AA\ feature, allowing more precise $\log({\rm Th}/{\rm Eu})$ measurements and reducing the associated statistical age uncertainties \citep[see, e.g.][]{Kordopatis2023A&A...674A.104K}\footnote{\url{https://line-detector.oca.eu}}. An additional possibility is offered by the weaker and less commonly studied Th~{\sc ii} 5989~\AA\ line, whose reliable measurement also requires very high spectral resolution and signal-to-noise. Although less widely used than the 4019~\AA\ feature, this line is receiving increasing attention in recent Th abundance studies \citep{ Azhari2025A&A...699A.276A, Furutsuka2026arXiv260913880F}.

Second, coeval stellar populations provide statistical leverage that is not available for individual field stars. In globular clusters, measurements of multiple stars can be treated as estimates of the same underlying formation epoch. If the star-to-star dispersion is dominated by independent random uncertainties, the statistical uncertainty on the mean decreases approximately as $1/\sqrt{N}$. At the same time, comparison of the observed dispersion with that expected from the individual uncertainties provides a means of identifying excess scatter associated with underestimated errors, star-dependent systematic effects, or possible intrinsic variations in the r-process abundance pattern. Systematic uncertainties shared by all cluster members, however, will not decrease with increasing sample size and will ultimately limit the {\bf accuracy of the absolute age determination}. Extending this approach to several globular clusters could provide {\bf more precise estimates of their relative formation times and, provided that the uncertainties in the initial Th/Eu production ratio are adequately constrained, more robust absolute ages for the oldest Galactic stellar populations. In combination with independent age indicators, these measurements could consequently contribute to establishing} a lower bound on the age of the Universe \citep[see, e.g.][]{Ludwig2010A&A...509A..84L,Massari2023A&A...680A..20M,Valcin2020JCAP...12..002V,Verde2024ARA&A..62..287V}.

Instruments combining very high spectral resolution, high throughput in the blue spectral region, and multiplex capabilities are needed to  make it possible to exploit these two complementary advantages simultaneously. By improving the precision of individual Th and Eu measurements while enabling observations of tens of stars within the same coeval population, such facilities could move nucleocosmochronology from an approach based primarily on individual stellar measurements towards a population-based framework. This would allow more precise mean ages to be derived while simultaneously testing the observational and nucleosynthetic assumptions underlying the Th/Eu chronometer.

 Combining radioactive nucleocosmochronology with independent age indicators, such as isochrone fitting and beryllium abundances \citep[see, e.g.][]{Pasquini2004A&A...426..651P,Pasquini2007A&A...464..601P}, offers a promising approach to obtaining more robust age constraints for the oldest stellar populations. Statistical frameworks such as Bayesian analysis could ultimately provide a natural way to combine these complementary constraints while accounting for their different sources of observational and systematic uncertainty.

\begin{acknowledgements}
L.M., R.E.G, S.R., M.T. acknowledge support from INAF through the Large Grants EPOCH, funding for the WEAVE project, the Mini-Grants Checs (1.05.23.04.02), and financial support under the National Recovery and Resilience Plan (PNRR), Mission 4, Component 2, Investment 1.1, Call for tender No. 104 published on 2 February 2022 by the Italian Ministry of University and Research (MUR), funded by the European Union – NextGenerationEU, through the Project ‘Cosmic POT’ (Grant Assignment Decree No. 2022X4TM3H, MUR). 
RS acknowledges support from the National Science Centre, Poland, research grant  2022/47/I/ST9/02358.
F.R. is a fellow of the Alexander von Humboldt Foundation, and acknowledges support by the Klaus Tschira Foundation. F.R. also acknowledges INAF for the 1.05.24.07.02 Mini grant 2024
“GALoMS–Galactic Archaeology for Low Mass Stars” (PI: C.T. Nguyen). Use was made of the Simbad database, operated at the CDS, Strasbourg, France, and of NASA’s Astrophysics Data System Bibliographic Services.  This research used Astropy,\footnote{http://www.astropy.org} a community-developed core Python package for Astronomy \citep{astropy:2018}. This work presents results from the European Space Agency (ESA) space mission Gaia. Gaia data are processed by the Gaia Data Processing and Analysis Consortium (DPAC). Funding for the DPAC is provided by national institutions, in particular the institutions participating in the Gaia MultiLateral Agreement (MLA). The Gaia mission website is \url{https://www.cosmos.esa.int/gaia}. The Gaia archive website is \url{https://archives.esac.esa.int/gaia}.
\end{acknowledgements}

\bibliographystyle{aa.bst}

\bibliography{biblio} 

\newpage

\begin{appendix} 

\section{Surface gravity from distance}
\label{sec:logg}

To derive the surface gravity, we used the fundamental relation

\begin{equation}
    \label{eq:fundamental}
    \mathrm{log}\,g = \mathrm{log} \,g_{\odot} + \mathrm{log} \frac{M}{M_\odot} + 4 \, \mathrm{log} \frac{T_{\mathrm{eff}}}{T_{\mathrm{eff} \odot}} +0.4 (M_{\mathrm{bol}} - M_{\mathrm{bol} \odot})
\end{equation}

\noindent where $M$ is the stellar mass and $M_{\rm bol}$ is the bolometric magnitude.
We assume $0.80M_\odot$ and the solar\footnote{IAU 2015 RESOLUTION B3} parameters \teff$_\odot = 5772$~K, \logg $= 4.44$~dex, and $M_{bol \odot} = 4.74$  \citep{prsa2016AJ....152...41P}.
$M_{\rm bol}$ was determined 
from the Gaia and 2MASS \citep{2mass2003tmc..book.....C}
magnitudes $BP$, $G$, $RP$, $J$, $H$, and $K_S$
with the python script {\it bcutil.py}\footnote{https://github.com/casaluca/bolometric-corrections}from \cite{casagrande2018MNRAS.479L.102C}.
Magnitudes were corrected for extinction using the reddening values in \cite{giribaldi2023A&A...679A.110G}, which were transformed 
into the Gaia and 2MASS systems using
the extinction law of \cite{fitzpatrick1999PASP..111...63F} as normalised per \cite{Schlafly2011ApJ...737..103S}\footnote{Using the code {\it Colte} \citep{casagrande2021} at \url{https://github.com/casaluca/colte}.}.

\section{Line fitting, additional abundances and uncertainties on Th abundances}
\label{sec:th_profiles_err}

In this Section, we present the spectral modelling of the Th feature at $\lambda$4019.129~\AA\ for the stars analysed in this work and our evaluation of the source of uncertainties on Th abundances. 

\begin{figure*}
    \centering
    \includegraphics[width=0.45\linewidth]{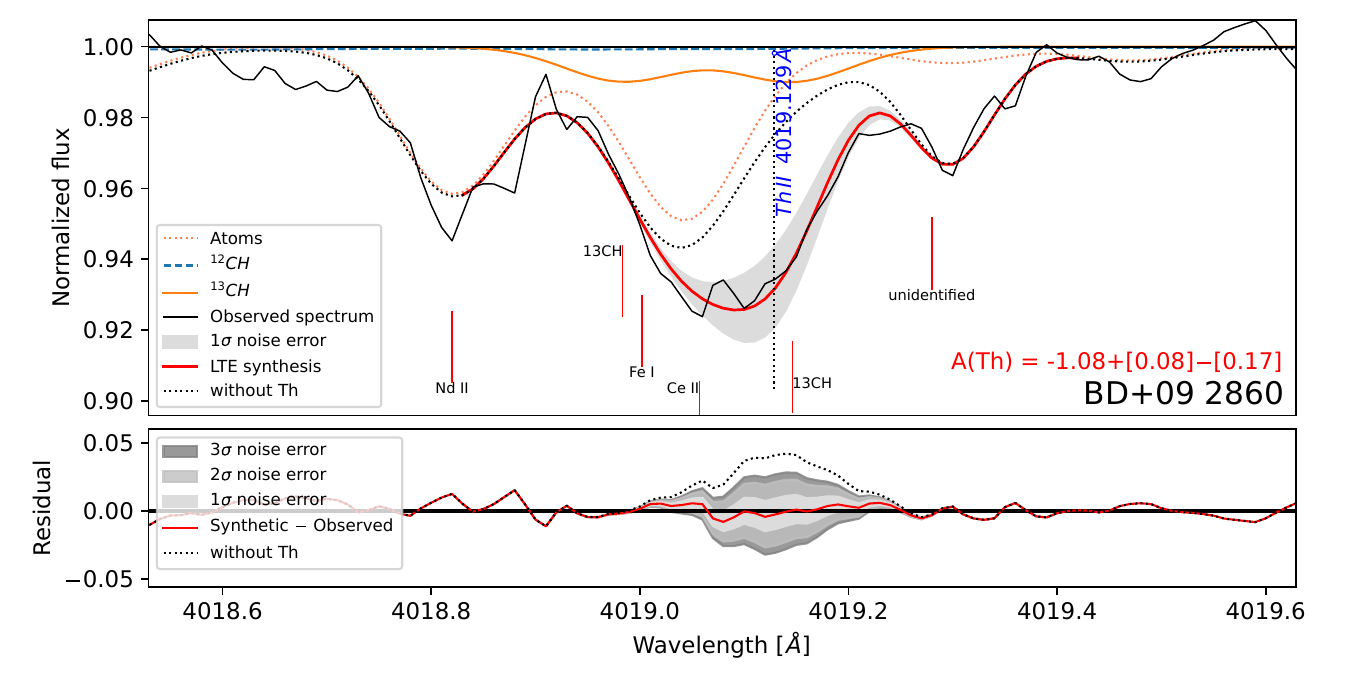}
    \includegraphics[width=0.45\linewidth]{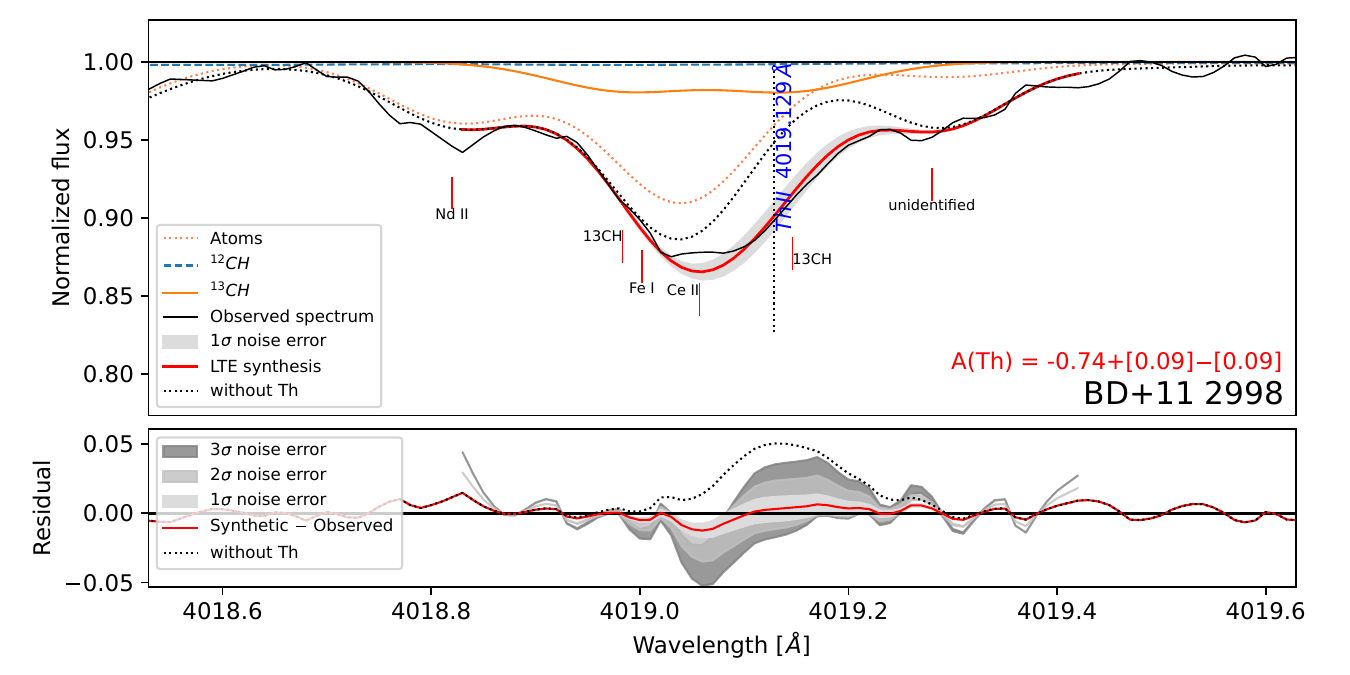}
    \includegraphics[width=0.45\linewidth]{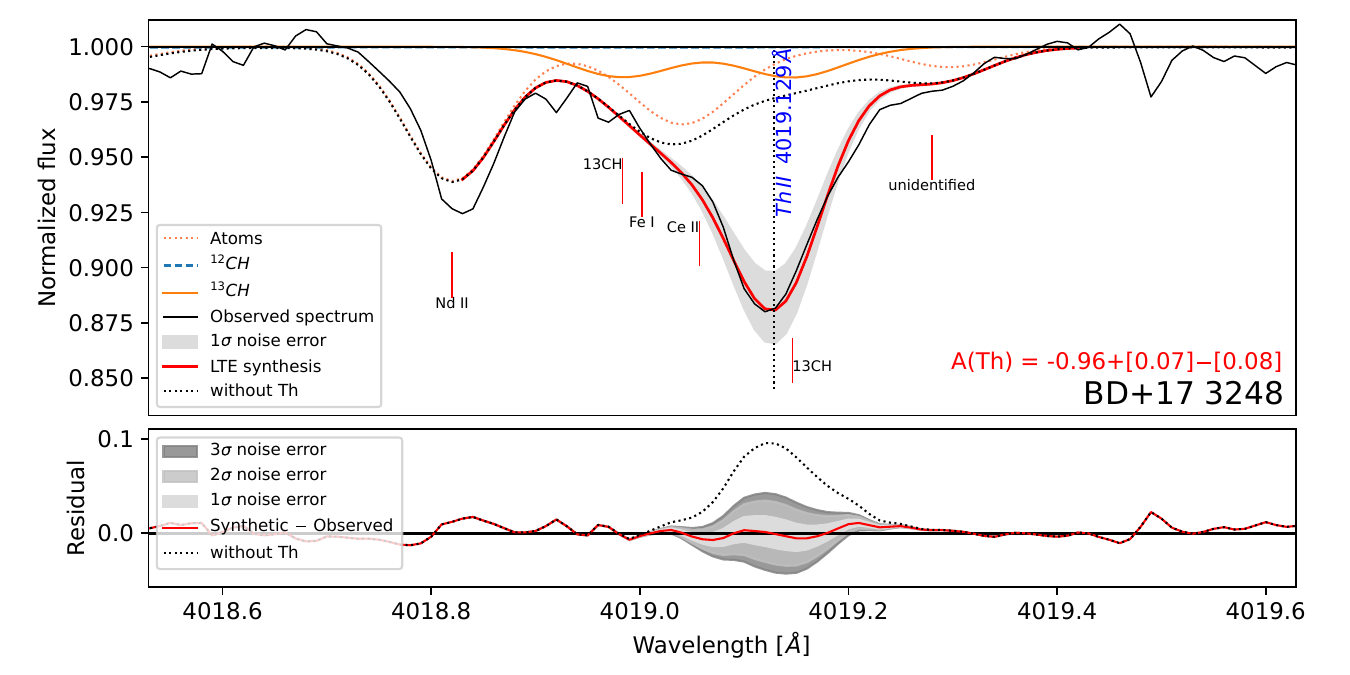}
    \includegraphics[width=0.45\linewidth]{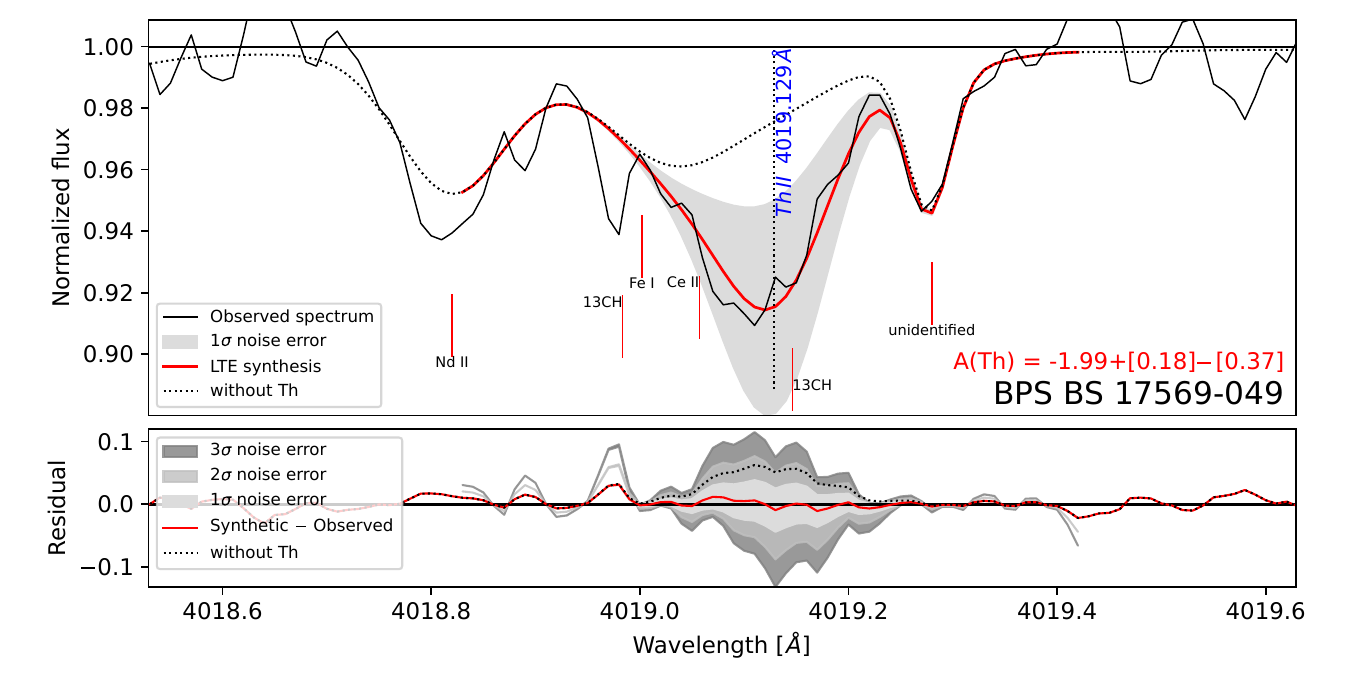}
    \includegraphics[width=0.45\linewidth]{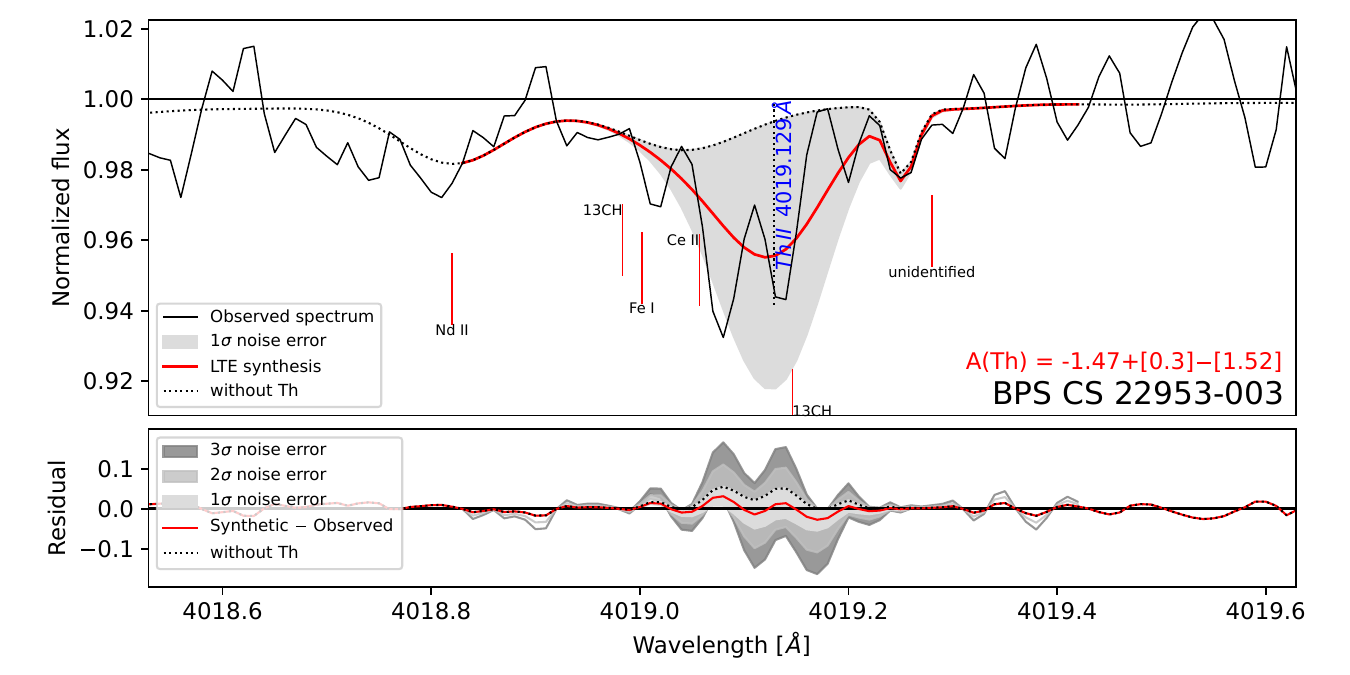}
    \includegraphics[width=0.45\linewidth]{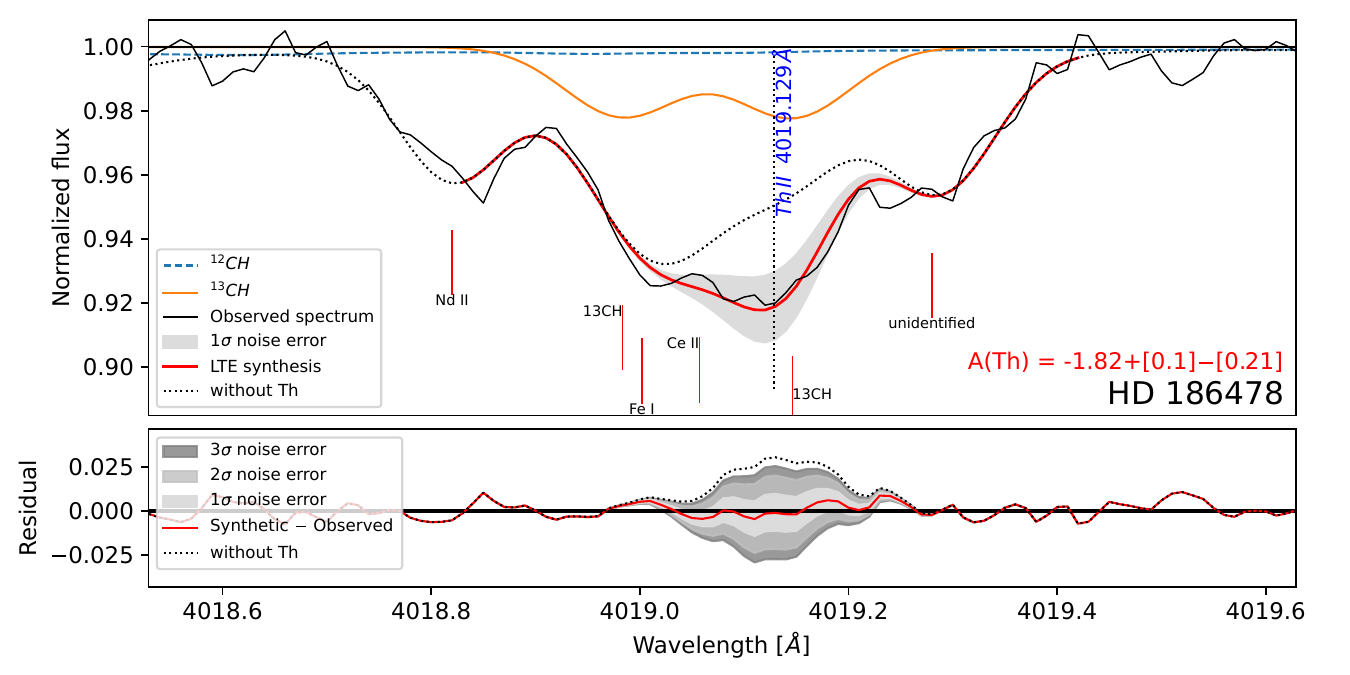}
    \includegraphics[width=0.45\linewidth]{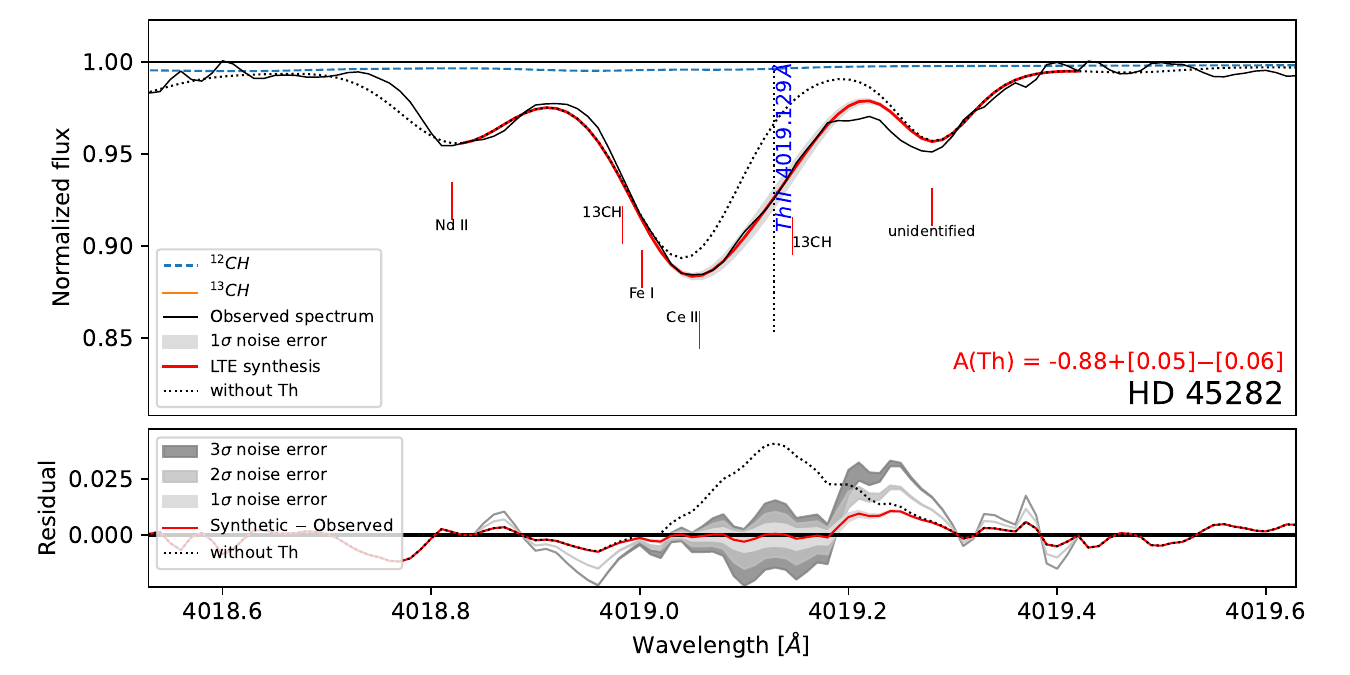}
    \includegraphics[width=0.45\linewidth]{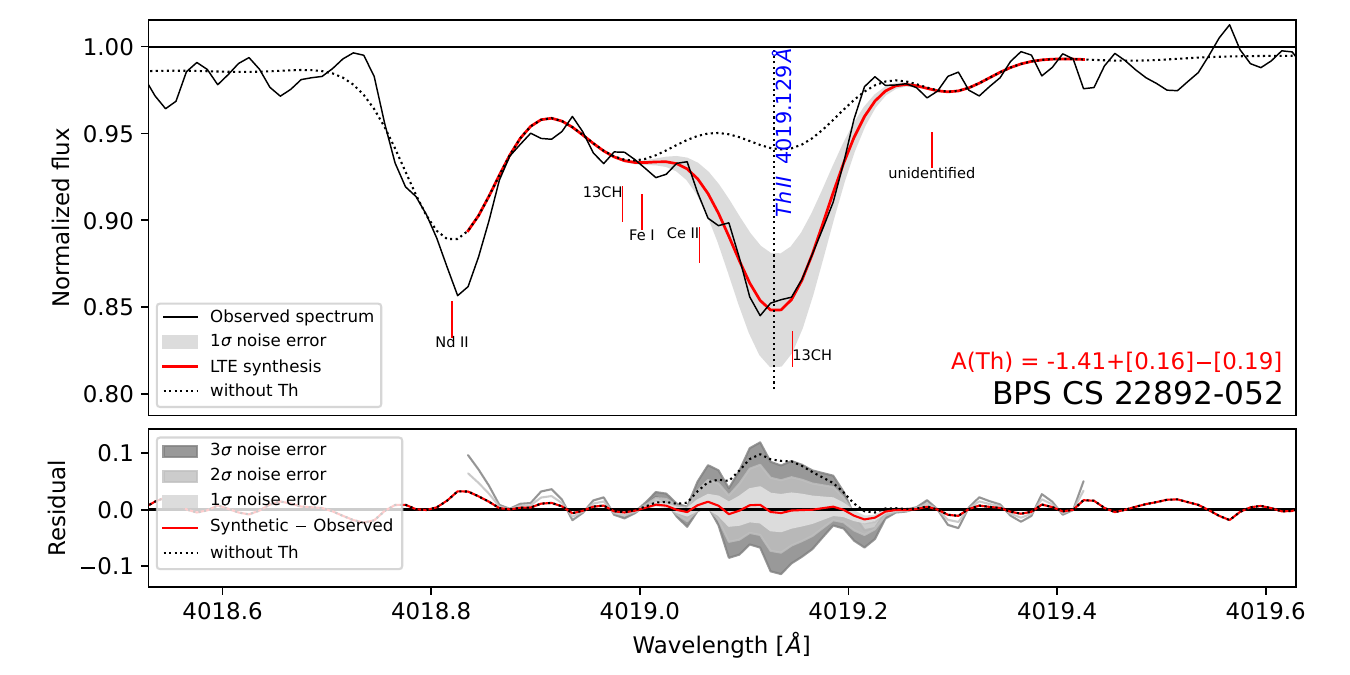}
    \caption{\tiny Modelling of the  thorium feature at $\lambda$4019.129~\AA. 
    Observational spectra are represented by the black lines.
    The Th line core is indicated by the dotted vertical line. Lines of species with expected significant contributions to the feature are marked in the plot.
    The abundance related to the best fit  is noted at the right bottom corner.
    Synthetic spectra without Th are represented by the dotted black lines.
    Residual panels display 1, 2, and 3$\sigma$ flux errors according to the legends.}
    \label{fig:plots}
\end{figure*}

\paragraph{Line fitting and   additional abundances.}  For the spectral modelling, the line at $\lambda$4019.129~\AA\ represents the main diagnostic for deriving Th abundances in metal-poor stars, although its measurement is challenging due to the presence of several blending features from neighbouring atomic species. To properly account for these contributions, synthetic spectra include all relevant blending species, whose measured abundances and isotopic ratios are reported in Table~\ref{tab:abundances}. The observed spectra are compared with synthetic spectra computed with and without Th, allowing us to assess the contribution of the Th line to the observed absorption feature and to derive the best-fit Th abundances. The resulting fits and residuals are shown in Fig.~\ref{fig:plots}. 

\paragraph{Uncertainties on thorium abundance}

\begin{figure*}
    \centering
    \includegraphics[width=0.24\linewidth]{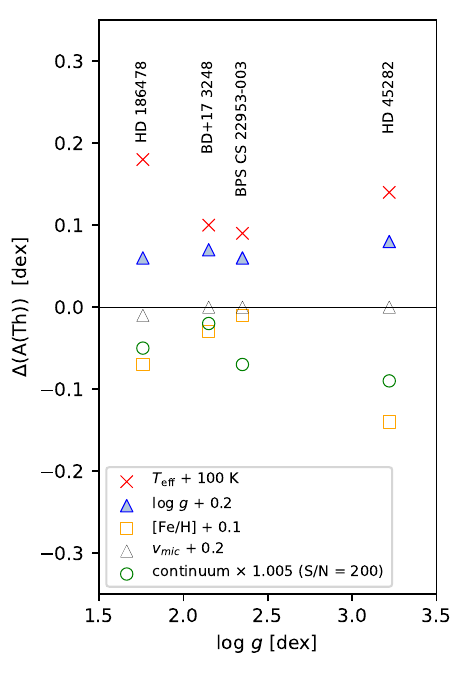}
    \includegraphics[width=0.24\linewidth]{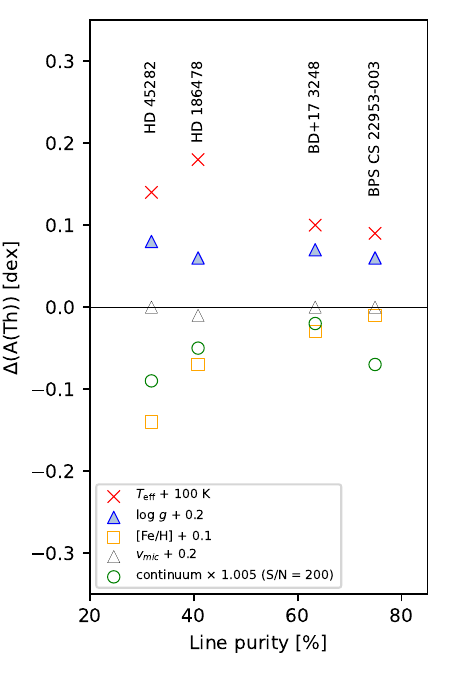}
    \includegraphics[width=0.24\linewidth]{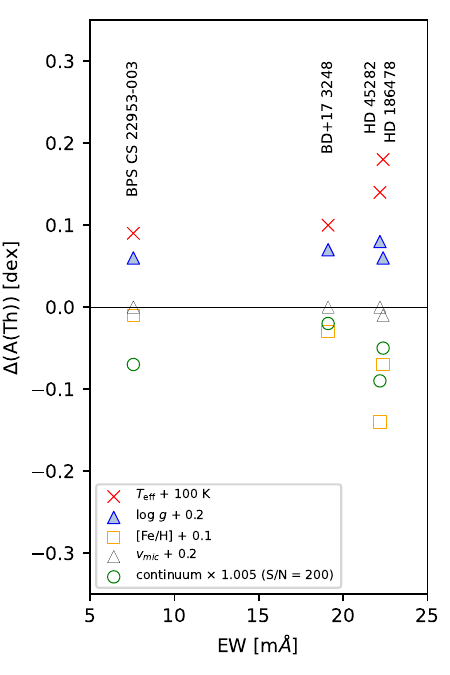}
    \includegraphics[width=0.24\linewidth]{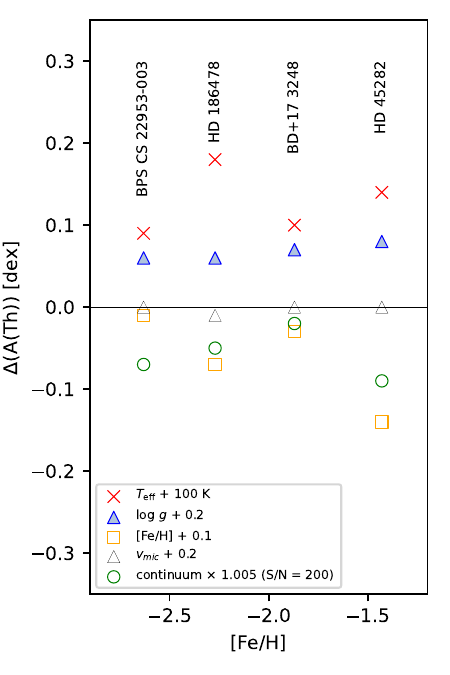}
    \caption{Variations of A(Th) induced by other parameter variations. The Th variations are plotted as function of several parameters in each plot. The parameter variations are indicated in the legends, and reflect typical uncertainties in the literature.}
    \label{fig:err_th}
\end{figure*}

The utility of Th abundances for stellar-age determinations ultimately depends on achieving sufficiently high overall precision. We therefore assessed the sensitivity of the derived Th abundances to uncertainties in the stellar parameters by considering representative stars from our sample that span different combinations of \teff, \logg, and [Fe/H] along both the upper and lower red giant branches (RGB).
Figure~\ref{fig:err_th} illustrates the individual contributions to the Th abundance uncertainty, which are discussed below. For each star, the total uncertainty in Th was then obtained by interpolating the corresponding contributions over the range of parameter uncertainties reported in Table~\ref{tab:parameters}.

In what follows, we summarise the impact of stellar parameters on Th abundances: 
\begin{description}
\item[$v_{\rm mic}$:] The impact is negligible.

\item[$\logg$:] The impact is moderate and nearly constant, with variations of approximately $\pm0.08$~dex in A(Th) for a $\pm0.20$~dex change in \logg.

\item[{\rm [Fe/H]}:] The impact depends on the purity of the Th line, defined as the ratio between the equivalent width (EW) of the Th line and that of the full feature within 0.2~\AA\ of the Th line centre. It decreases with increasing line purity, becoming marginal for purities larger than $\sim$60\%.

\item[Continuum placement:] The impact depends on both the line purity and EW, being stronger for weaker and less-pure features.

\item[\teff:] For a typical uncertainty of $\sim$100~K\footnote{This includes the internal uncertainty of the method (50--100~K) and the accuracy of the \teff\ scale. For modern studies calibrated against the Gaia Benchmarks \citep{heiter2015A&A...582A..49H} and the \titans, the latter is approximately 50~K.}, the impact is in most cases the largest among the considered parameters. It also increases towards lower \teff, i.e. as the star approaches the tip of the RGB. This highlights the importance of having well-characterised reference stars, such as the \titans, for which \teff\ can be determined with the highest possible precision.
\end{description}

Aside from the uncertainties of the parameters associated to the physical properties of the stars, the most relevant factor limiting the precision of the Th abundance, and thus of the stellar age via nucleochronology, is the quality of the spectral and atomic data. 
Spectral resolution, spectral sampling, and S/N, govern the maximum precision to which A(Th) can be measured. 
As the quantities in Table~\ref{tab:parameters} show, for the UVES configuration (sampling of $\approx$0.015 \AA/pixel) at the resolution $R \sim 40\,000$, at least S/N = 250 is required to obtain a line fit precision of $\approx$0.1~dex for a star with normal [Th/Fe] $\sim 0.5$~dex.
To achieve a reasonable age precision of 2.5~Gyr, the combined precision of Th and Eu abundances must be of $\sim$0.05~dex, as it will be shown in Sect.~\ref{sec:ages} and Fig.~\ref{fig:sensitivity}.
Therefore, the total uncertainty of A(Th) cannot be larger than $\approx$0.04~dex (leaving room for a similar, or even lower,  uncertainty value for A(Eu)). 

At a fixed spectral resolution $R = 40,000$, improving the precision of the Th abundance measurement ultimately requires very high S/N, which can be prohibitively expensive in exposure time and is therefore often unattainable for individual stars. A substantial improvement can instead be achieved by increasing the spectral resolution. For example, at $R = 80,000$, an A(Th) precision of 0.04~dex can be obtained at S/N $\sim 350$, compared to S/N $\approx 560$ with the current UVES configuration. This is particularly relevant in the context of multi-object observations of metal-poor star clusters, where multiple twin stars can be observed simultaneously and their spectra co-added. For example, observing four twin stars at S/N $\approx 200$ would yield a combined S/N $\approx 400$, enabling A(Th) line-fitting precisions of $\approx 0.03$~dex with substantially reduced exposure times.


\begin{table*}
\caption{Additional abundances}
\label{tab:abundances}
\centering
\tiny 
\begin{threeparttable}
\begin{tabular}{l|cccccccccc}
\hline\hline
Star & A(C) & $^{12}$C/$^{13}$C &  A(Ni) & A(Ce) & A(Co)\\
 &  [dex] & & [dex] & [dex] & [dex] \\
\hline
BD+09 2860 & $6.33 \pm 0.02$ & 95 & $4.49 \pm 0.03$ & $0.05 \pm 0.02$ & ---\\ 
BD+11 2998 & $7.25\pm 0.02$ & 96 & $5.00 \pm 0.04$ & $0.40 \pm 0.01$ & $3.73 \pm 0.05$\\
BD+17 3248 & $6.09 \pm 0.05$ & 82 & $4.21 \pm 0.03$ & $0.03 \pm 0.02$ & ---\\
BPS BS 17569-049  & $5.71\pm 0.03$ & 91 & $3.40 \pm 0.05$ & $-0.98 \pm 0.04$ & --- \\
BPS CS 22953-003 & $6.22\pm 0.03$ & >100 & $3.17 \pm 0.04$ & $-0.65 \pm 0.05$ & ---\\
HD 186478 & $5.71\pm0.02$ & 89 & $3.71 \pm 0.06$ & $-0.62 \pm 0.05$ & $2.52\pm 0.03$\\
HD 45282 & $6.87 \pm 0.02$ & >100 & $4.64 \pm 0.05$ & $0.27\pm0.11$ & $3.48\pm0.10$ \\
BPS CS 22892-052 & $6.64 \pm 0.03$ & 0.92 & --- & $-0.22 \pm  0.02$ & --- \\
\hline
Star & A(Eu) & A(Eu) &  A(Eu) & A(Eu) & A(Eu) & A(Eu)\\
 &  4129 (LTE) & 4129 (NLTE) & 4205 (LTE) & 4205 (NLTE) & 6437 (LTE) & 6645 (LTE) \\
\hline
BD+09 2860 & $-0.72$ & $-0.65$ & $-0.63$ & $-0.56$ & $-0.50$ & $-0.57$ \\ 
BD+11 2998 & $-0.40$ & $-0.33$ & $-0.28$ & $-0.21$ & $-0.27$ & --- \\
BD+17 3248 & $-0.51$ & $-0.45$ & $-0.48$ & $-0.42$ & $-0.46$ & $-0.39$ \\
BPS BS 17569-049 & $-1.65$ & $-1.57$ & $-1.66$ & $-1.57$ & --- & $-1.56$ \\
BPS CS 22953-003 & $-1.16$ & $-1.10$ & $-1.16$ &  $-1.09$ & --- & ---- \\
HD 186478 & $-1.31$ & $-1.26$ & $-1.27$ & $-1.21$ & --- & --- \\
HD 45282 &  $-0.50$ & --- & $-0.40$ & --- & $-0.39$ & $-0.32$ \\
BPS CS 22892-052 & $-0.79$ & --- & $-0.78$ & --- & --- & $-0.78$\\
\hline
\end{tabular}
\begin{tablenotes}
\item{} \textbf{Notes.} {Uncertainties represent the dispersion of the measurements of the lines. The second part of the table lists individual A(Eu) from each line  as indicated in the header (in \AA). Non-LTE models were not available in our grids for HD~45282 and BPS CS 22892-052. Noisy lines at 6437~\AA\ prevented abundance measurements for some stars.} 
\end{tablenotes}
\end{threeparttable}
\end{table*}

\begin{figure*}
    \centering
\includegraphics[width=0.9\linewidth]{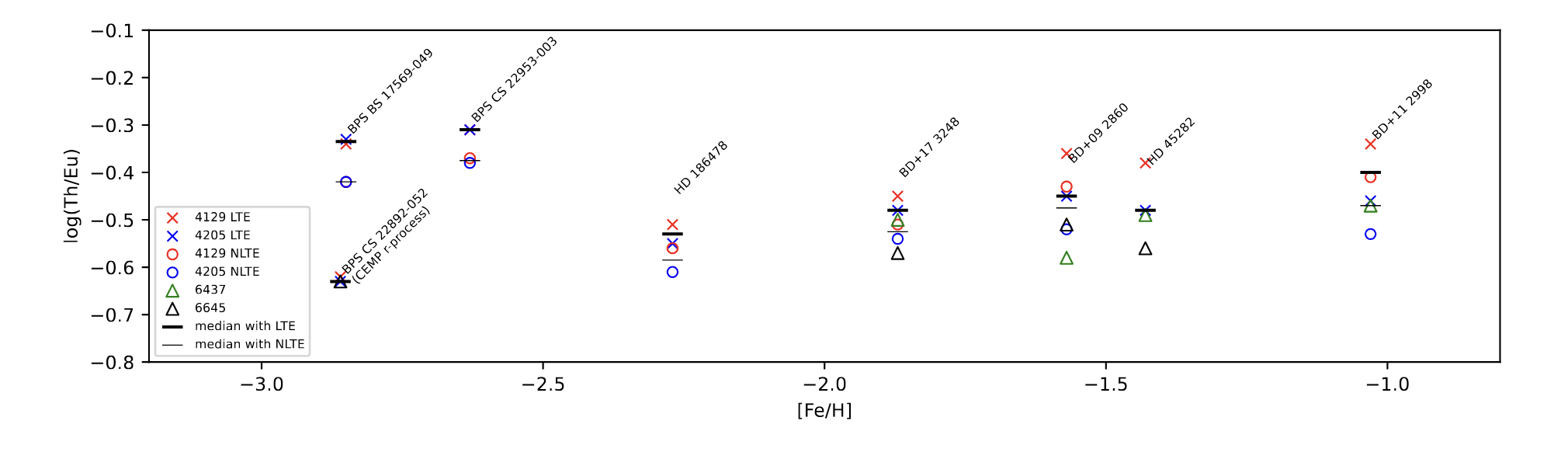}
    \caption{log(Th/Eu) abundance as function of [Fe/H]. Symbols represent quantities computed considering A(Eu) from each line as indicated in the legends. Median considering LTE and NLTE A(Eu) values for each star are shown by black thick and thin dashes, respectively. }
    \label{fig:theu_comp}
\end{figure*}

 \end{appendix}


\end{document}